\documentclass[a4paper,fleqn,usenatbib]{mnras}

\usepackage{newtxtext,newtxmath}

\usepackage[T1]{fontenc}
\usepackage{ae,aecompl}
\usepackage{etoolbox}
\makeatletter
\patchcmd\@combinedblfloats{\box\@outputbox}{\unvbox\@outputbox}{}{%
   \errmessage{\noexpand\@combinedblfloats could not be patched}%
}%
 \makeatother

\usepackage{graphicx}	
\usepackage{amsmath}	
\usepackage{amssymb}	
\usepackage{comment}

\newcommand {\kms} {\,{\rm km\,s}^{-1}}

\newcommand {\mo}{{\rm M}_\odot}

\newcommand {\moyr}{\,{\rm M_\odot\,\rm yr}^{-1}}

\def\hi{\ifmmode{\rm HI}\else{H\/{\sc i}}\fi} 
\def\ovii{\ifmmode{\rm OVII}\else{O\/{\sc vii}}\fi} 
\def\oviii{\ifmmode{\rm OVIII}\else{O\/{\sc viii}}\fi}

\newcommand{\aref}[1]{\hyperref[#1]{Appendix~\ref{#1}}}

\title[Life cycle of the CMZ]{The Life Cycle of the Central Molecular Zone. I: Inflow, Star Formation, and Winds}

\author[L. Armillotta et al.]{Lucia Armillotta$^{1}$\thanks{E-mail: lucia.armillotta@anu.edu.au}, Mark R. Krumholz$^{1,2}$,  Enrico M. Di Teodoro$^{1}$,   and \newauthor N. M. McClure-Griffiths$^{1}$\\
$^{1}$Research School of Astronomy and Astrophysics - The Australian National University, Canberra, ACT, 2611, Australia\\
$^{2}$Centre of Excellence for Astronomy in Three Dimensions (ASTRO-3D), Australia}
\date{Accepted XXX. Received YYY; in original form ZZZ}

\pubyear{2018}

\begin{document}
\label{firstpage}
\pagerange{\pageref{firstpage}--\pageref{lastpage}}
\maketitle

\begin{abstract}
We present a study of the gas cycle and star formation history in the central 500 pc of the Milky Way, known as Central Molecular Zone (CMZ). Through hydrodynamical simulations of the inner 4.5 kpc of our Galaxy, we follow the gas cycle in a completely self-consistent way, starting from gas radial inflow due to the Galactic bar, the channelling of this gas into a dense, star-forming ring/stream at $\approx200-300$~pc from the Galactic centre, and the launching of galactic outflows powered by stellar feedback. We find that star formation activity in the CMZ goes through oscillatory burst/quench cycles, with a period of tens to hundreds of Myr, characterised by roughly constant gas mass but order-of-magnitude level variations in the star formation rate. Comparison with the observed present-day star formation rate of the CMZ suggests that we are currently near a minimum of this cycle. Stellar feedback drives a mainly two-phase wind off the Galactic disc. The warm phase dominates the mass flux, and carries $100-200\%$ of the gas mass converted into stars. However, most of this gas goes into a fountain and falls back onto the disc rather than escaping the Galaxy. The hot phase carries most of the energy, with a time-averaged energy outflow rate of $10-20\%$ of the supernova energy budget. 
\end{abstract}

\begin{keywords}
hydrodynamics - methods: numerical - Galaxy: centre - Galaxy: evolution - galaxies: star formation
\end{keywords}



\section{Introduction}
\label{Introduction}

The innermost region of the Milky Way - within a radius of 500~pc - represents an extreme environment in our Galaxy. This region, known as Central Molecular Zone \citep[CMZ;][]{Morris&Serabyn96}, contains a large reservoir of molecular gas \citep[$M_\mathrm{CMZ}\sim 3-7 \times10^7\,\mo$, e.g.][]{Ferriere+07, Molinari+11, Longmore+13a}, characterised by volume and surface gas densities $1-4$ orders of magnitude larger than those measured in the solar neighbourhood and in the outer regions of the Galactic disc \citep[$n_\mathrm{CMZ} \sim 10^4\,\rm{cm}^{-3}$, $\Sigma_\mathrm{CMZ}>10^2\,\mo\,\rm{pc}^{-2}$; e.g.][]{Kruijssen&Longmore13, Kruijssen+14, Ginsburg+16, Battersby+17}. Most of this gas is located in a stream, or a partially filled ring, at $\sim 100-150$~pc from the Galactic centre, presenting a highly asymmetric distribution \citep[e.g.][]{Molinari+11, Kruijssen+15, Henshaw+16}. Within this stream there are a number of molecular clouds, from actively star-forming (e.g. Sgr~A, Sgr~B, Sgr~C) to almost starless (e.g. G0.253+0.016, known as The Brick).

Despite the high gas densities, the present-day star formation rate (SFR) of the CMZ \citep[$\sim 0.04-0.1 \moyr$, e.g.][]{Yusef-Zadeh+09, Immer+12, Longmore+13a, Barnes+17} is about one order of magnitude lower than the SFR of gas of comparable volume density that is not in the CMZ.
Deviations from the star formation patterns observed at larger radii are common for the nuclear regions of star-forming galaxies. On average, nuclear regions present shorter depletion times (ratio of the gas surface density to the star formation surface density) and, therefore, more efficient star formation per unit molecular gas mass compared to disc regions. However, while the latter form stars with a fairly constant depletion time, the nuclear regions exhibit a very broad range of depletion times \citep[variations over $\sim 1$~dex,][]{Leroy+13, Utomo+17}. A possible explanation is that star formation is episodic in galactic centres and the CMZ might be at the minimum of a longer star formation period \citep{Sarzi+07, Kruijssen+14, Krumholz+15a, Krumholz+17}. In this scenario, nuclear regions are characterized by a dynamical time (which regulates collapse and star formation) shorter than the stellar evolution time (which regulates feedback). Thus they keep forming stars until a critical point when feedback becomes effective, leading to alternate cycles of burst/quenching \citep[see also simulations by][]{Torrey+17}.

Clearly, a proper understanding of the star formation cycle in the CMZ requires accounting for the gas dynamics in the central regions of our Galaxy. It is now well established that the Milky Way is a barred galaxy and gas dynamics is strongly influenced by the presence of a non-axisymmetric gravitational potential. The idea proposed by \citet{Binney+91} (see also \citealt{Kim&Stone12, Sormani+15, Li+16, Sormani+17}) is that gas in the outer parts of the bar slowly drifts towards the Galactic centre following a sequence of $x_1$ orbits, which are closed orbits elongated parallel to the bar major axis. These orbits are more and more elongated as the Galactic centre is approached, until becoming self-intersecting. At this point, gas is shocked towards $x_2$ orbits, which are closed orbits elongated parallel to the bar minor axis. The transition happens through dust lanes that carry gas from the $x_1$ orbits to the $x_2$ orbits. Recently, \citet{Sormani+19} have modelled $^{12}$CO data of the inner Galaxy to determine the mass inflow rate along the dust lanes towards the CMZ. They estimated a total inflow of $\sim 2.7\moyr$, thus showing that only a few percent of inflowing gas is converted into stars. It appears that most of the gas is eventually expelled from the CMZ through large-scale outflows.

The energetics of a galactic-scale outflow from the CMZ represents an important and puzzling problem in itself. The most compelling evidence for the existence of a galactic outflow in the centre of the Milky Way comes from the so-called Fermi Bubbles, two giant lobes extending up to $\sim 10$~kpc above and below the Galactic centre associated with non-thermal emission \citep[e.g.][]{Bland-Hawthorn&Cohen03, Su+10}. Whether these relativistic particles are the result of AGN activity, or whether they are cosmic rays produced by star formation is matter of debate. In this paper we do not address this question, since we do not include cosmic ray production or transport in our study. In addition to the Fermi Bubbles, recent observations have shown the presence of neutral \citep{McClure-Griffiths+13, DiTeodoro+18} and ionised \citep[e.g.][]{Fox+15} warm outflowing gas a few kpc above the plane in the Galactic centre region. The association between this gas and the Fermi Bubbles has not been fully established yet. \citet{DiTeodoro+18} argued that cold/warm gas might be entrained in a hot wind powered by the star formation activity in the CMZ. However, this conclusion is based on an idealized model and it is not clear whether the observed energetics and mass of the warm outflow are compatible with stellar feedback from the CMZ.

The aim of our project is to interpret and connect all these observational features characterizing the central region of our Galaxy and, particularly, the CMZ. We perform a three-dimensional hydrodynamical simulation of the inner 4.5~kpc of the Milky Way in order to follow the gas cycle in this region, starting from the gas radial inflow due to the Galactic bar.
Our work includes a combination of features absent from previous treatments of the problem: a realistic gravitational potential including contributions from dark matter, the stellar disc, the bulge, the Galactic bar, and gas self-gravity, a robust treatment of star formation and feedback (including stochastic supernovae), and high mass and spatial resolution that allows us to resolve all gas phases from cold molecular material at $T \sim 20$~K to shock-heated supernova remnants at $T\sim10^8$~K. In this paper we present our simulation methodology and a top-level view of the outcome, focusing on the budget of mass inflow, star formation, and galactic wind. Subsequent papers in this series will present details of the observational post-processing and detailed comparisons with observations, and will compare the Milky Way's CMZ to analogous regions in other nearby galaxies.
The results of this study may have implications in the broader context of galaxy evolution, as the extreme properties of the CMZ are similar to those observed in starburst nuclei of nearby galaxies and high-redshift galaxies at the peak of their star formation history. 

The paper is organized as follows. In \autoref{Method}, we introduce the initial conditions of our simulation and we briefly describe the main features of the code. In \autoref{Results}, we present an analysis of the simulation outcomes. In \autoref{Discussion}, we discuss our work in relation to observational findings and  other computational works. Finally, in \autoref{Conclusions}, we summarise our main results. 

\section{Method}
\label{Method}

\subsection{Numerical methods}
\label{Methods}

The simulation described in this work is run with \textsc{GIZMO} \citep{Hopkins15}, a parallel magneto-hydrodynamical code, based on a mesh-free, Lagrangian finite-volume Godunov method designed to capture advantages of both grid-based and particle-based methods. We use the Meshless Finite Mass solver on a reflecting-boundary domain and assume gas to follow an ideal equation of state with a constant adiabatic index $\gamma = 5/3$. Gravity is solved by the Tree method solver described in \citet{Springel05}, with the minimum gravitational softening length, $s$, for both star and gas particles equal to 1~pc. \footnote{In subsequent papers in this series we will analyse re-simulations of parts of this evolution at higher resolution and with smaller softening length, but in this paper we limit ourselves to our long-term global simulations with $s=1$~pc.}.

For this study, we have included in the code an external gravitational force specific to our problem. Moreover, we have implemented algorithms for radiative cooling, star formation and stellar feedback. In the following, we describe the details of our implementation.

\subsubsection{External gravity}
\label{Potential}

\begin{figure}
\includegraphics[width=0.48\textwidth]{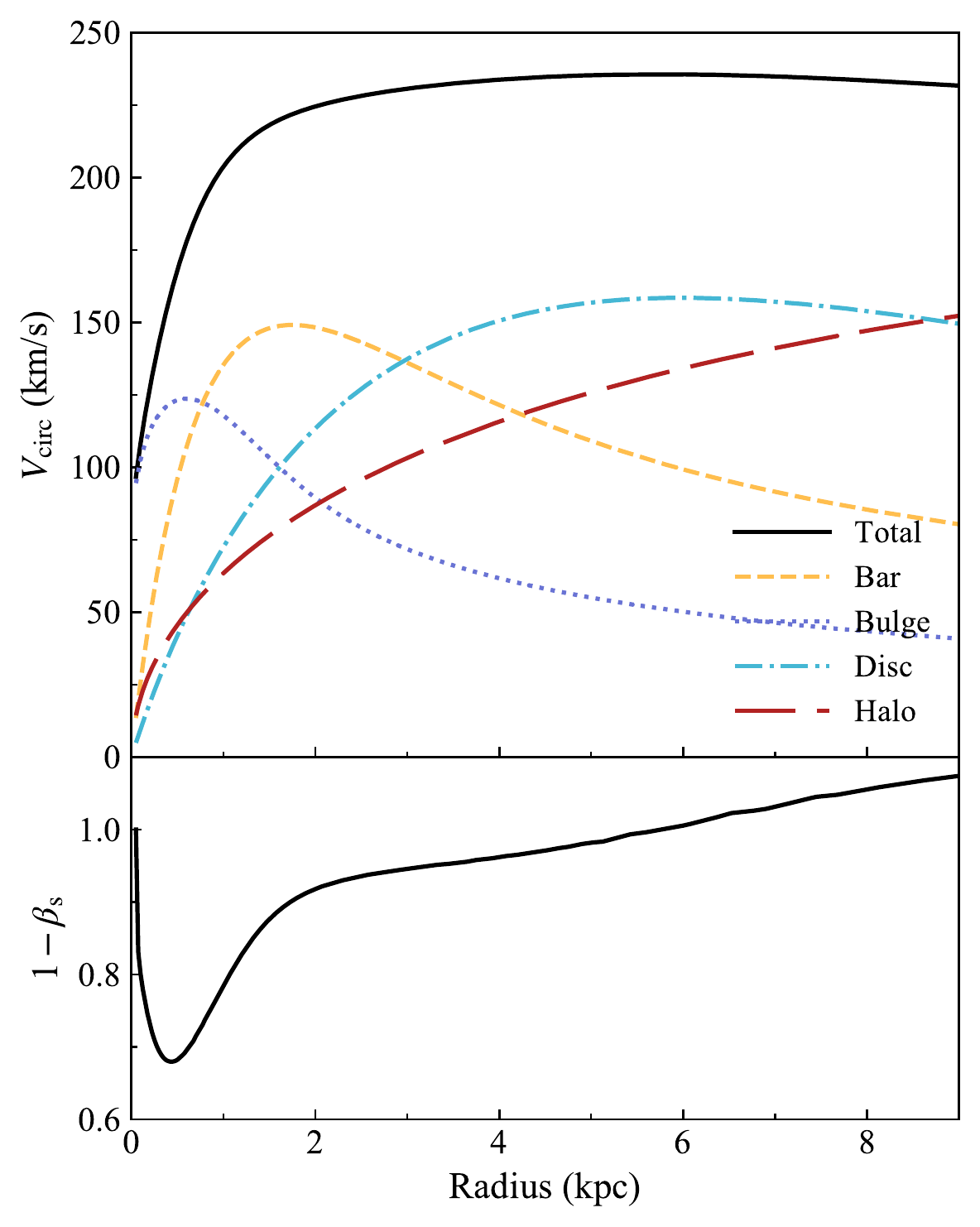}
\caption{Circular velocity curve induced by the gravitational potential used in this work and described in \autoref{Potential} (\textit{top panel}) and dimensionless shear parameter, $1-\beta_\mathrm{s}\equiv 1-\mathrm{d \,ln} \,V_\mathrm{c} (R) / \mathrm{d\, ln}\, R$ (\textit{bottom panel}; note that a value of unity for this parameter corresponds to a flat rotation curve, and a value of 0 corresponds to solid body rotation). The full black curve is the total velocity. Yellow dashed, blue dotted, cyan dot-dashed and red long-dashed lines denote the contributions of the bar, the bulge, the disc (thin and thick) and the halo, respectively.}
\label{Rotcur}
\end{figure}

For consistency with recent simulations of gas flows in the Galactic centre \citep{Ridley+17, Sormani+17}, we use the best-fit Milky Way potential by \citet{McMillan17} as modified by \citet{Ridley+17} to account for the non-axisymmetric stellar bar component.
The total potential is given by
\begin{equation}
\label{eq:pottotal}
 \phi_\mathrm{tot} = \phi_\mathrm{B} + \phi_\mathrm{b} + \phi_\mathrm{d} + \phi_\mathrm{h} \;,
\end{equation}
where $\phi_\mathrm{B}$, $\phi_\mathrm{b}$, $\phi_\mathrm{d}$, $\phi_\mathrm{h}$ are the contributions of the bar, the bulge, the disc (thick and thin) and the dark-matter halo, respectively.
The gravitational potential of each component is calculated through the integration of the Poisson equation \citep[see e.g. chapter 2 in][]{Binney&Tremaine08} using the publicly-available \textsc{Galpynamics} package\footnote{https://github.com/editeodoro/galpynamics}.
In the following, we summarize the density profiles that generate the potential of each component. We denote the cylindrical radius as $R^2=x^2+y^2$ and the spherical radius as $r^2=x^2+y^2+z^2=R^2+z^2$.

The density of the stellar bar is an exponential prolate ellipsoid \citep[e.g.][]{Wegg&Gerhard13},
\begin{equation}
\label{eq:bar}
\rho_\mathrm{B} (x,y,z) = \rho_\mathrm{B,0}\exp(-a/a_0) \;,
\end{equation}
where $\rho_\mathrm{B,0}$ is the central density, $a_0$ is the equivalent of a scale radius and $a^2 = x^2+(y^2+z^2)/q^2_\mathrm{B}$ with $q^2_\mathrm{B}$ being the axis ratio of the bar.
Consistent with \citet{Ridley+17}, we assume the following parameters for the MW bar: $\rho_\mathrm{B,0}=5\: \mathrm{M_\odot \, pc^{-3}}$, $a_0=0.75\: \mathrm{kpc}$ and $q_\mathrm{B}=0.5$. During the simulation, the bar rotates with a constant pattern speed $\Omega_\mathrm{p} = 40 \: \mathrm{km \, s^{-1} \, kpc^{-1}}$ \citep[e.g.][]{Wegg+15}. Inner and outer Lindblad resonances are located at $R_\mathrm{ILR} = 1.2 \: \mathrm{kpc}$ and $R_\mathrm{OLR}=9.6 \; \mathrm{kpc}$, respectively, while the co-rotation radius is at $R_\mathrm{CR} = 5.9 \: \mathrm{kpc}$. 

The bulge has an oblate, spheroidal truncated power-law density distribution \citep[e.g.][]{Bissantz&Gerhard02},
\begin{equation}
\label{eq:bulge}
\rho_\mathrm{b} (R,z) = \rho_\mathrm{b,0}\left( \frac{m}{m_\mathrm{b}}\right)^{-\alpha_\mathrm{b}}\exp(-m/r_\mathrm{b}) \;,
\end{equation}
where $m^2 = x^2+y^2+z^2/q^2_\mathrm{b} = R^2+z^2/q^2_\mathrm{b}$.
We use a central density $\rho_\mathrm{b,0} = 0.8 \: \mathrm{M_\odot \, pc^{-3}}$, a power-law $\alpha_\mathrm{b} = 1.7$, a truncation radius $r_\mathrm{b}= 1\: \mathrm{kpc}$, $m_\mathrm{b}= 1\: \mathrm{kpc}$ and axis ratio $q_\mathrm{b} = 0.5$.

The stellar disc includes a thin and a thick exponential component \citep[e.g.][]{Gilmore&Reid83},
\begin{equation}
\label{eq:stellardisk}
\rho_\mathrm{d,*} (R,z) = \frac{\Sigma_1}{2z_1}\exp\left( - \frac{\mid z \mid}{z_1} - \frac{R}{R_1} \right) + \frac{\Sigma_2}{2z_2}\exp\left( - \frac{\mid z \mid}{z_2} - \frac{R}{R_2} \right)\;,
\end{equation}
where $\Sigma_1$ and $\Sigma_2$ are the central surface densities of the thin and thick discs, respectively, $R_1$ and $R_2$ are their scale radii, and $z_1$ and $z_2$ are their scale heights.
The assumed thin-disc parameters are $\Sigma_1 = 850 \:  \mathrm{M_\odot \, pc^{-2}}$, $R_1=2.5 \; \mathrm{kpc}$, $z_1 = 0.3 \: \mathrm{kpc}$ and the thick-disc parameters are $\Sigma_2 = 174 \:  \mathrm{M_\odot \, pc^{-2}}$, $R_2=3.0\; \mathrm{kpc}$, $z_2 = 0.9 \: \mathrm{kpc}$.

Finally, the dark matter halo is assumed to follow a classic Navarro, Frenk and White profile \citep{Navarro+96},
\begin{equation}
\label{eq:gasdisk}
\rho_\mathrm{h} (r) =  \frac{\rho_\mathrm{0,h}}{(r/r_\mathrm{h})(1+r/r_\mathrm{h})^2}\;,
\end{equation}
with central density $\rho_\mathrm{0,h} = 8.11\times10^{-3} \: \mathrm{M_\odot \, pc^{-3}}$ and scale radius $r_\mathrm{h} = 19.6 \; \mathrm{kpc}$.

\autoref{Rotcur} shows the circular velocity curves resulting from the gravitational potentials described above. 
Rotation curves of the axisymmetric components (bulge, disc and halo) at $z=0$ are calculated simply as 
\begin{equation}
V^2_\mathrm{c} (R) = - R \dfrac{ \partial \phi}{ \partial R}\;,
\label{vrot}
\end{equation}
where $\phi = \phi_\mathrm{tot} - \phi_\mathrm{b}$ (see \autoref{eq:pottotal}). 
The velocity induced by the non-axisymmetric bar is calculated through a multipole expansion of the bar potential and considering only the symmetric monopole term. 
The total circular velocity is shown as a full black line in \autoref{Rotcur}. 
In this Milky Way potential, the Sun rotates at a velocity $V_\odot \simeq 235 \; \mathrm{km \, s^{-1}}$ at $R_\odot = 8.2 \: \mathrm{kpc}$. The bottom panel of \autoref{Rotcur} shows the dimensionless shear parameter, $1-\beta_\mathrm{s}$, where $\beta_\mathrm{s} \equiv \mathrm{d \,ln} \,V_\mathrm{c} (R) / \mathrm{d\, ln}\, R$. We see that our adopted potential has a shear minimum at  $\approx 400$~pc from the Galactic centre. Note that the location of the shear minimum for this potential differs from the $\approx 100$~pc produced by the empirically-determined \citet{Launhardt+02} potential adopted by \citet{Krumholz+15a} and \citet{Krumholz+17} for their models. 

We provide the values of the Milky Way gravitational potential to \textsc{GIZMO} through a look-up table depending on the Cartesian coordinates $x$, $y$, $z$. In this table, the gravitational potential is evaluated in the rest frame corotating with the bar. To estimate the gravitational potential in the simulation frame, we first calculate the gravitational accelerations along the three spatial directions in the time-dependent bar frame and then rotate these values in the simulation frame.

\subsubsection{Star formation}
\label{SF}

We parametrise the SFR in our simulation as
\begin{equation}
\dot {\rho}_\mathrm{SF} = \epsilon_\mathrm{ff} \dfrac{\rho_\mathrm{g}}{t_\mathrm{ff}}\;,
\label{rhoSF}
\end{equation}
where $\rho_\mathrm{g}$ is the local gas density, $t_\mathrm{ff} = \sqrt{{3 \pi}/{32 G \rho_\mathrm{g}}}$ is the local free-fall time and $\epsilon_\mathrm{ff}$ is the star formation efficiency.

The value of $\epsilon_\mathrm{ff}$ has been studied extensively using multiple observational methods, which yield a mean value $\epsilon_\mathrm{ff} \approx 0.01$ over a very wide range in gas density, with a scatter $\approx 0.3 - 0.5$~dex
(e.g.~\citealt{Krumholz&Tan07, Krumholz+12, Heyer+16, Vutisalchavakul+16, Utomo+18}; for a compilation of additional observations, as well as a discussion of contrasting results such as those of \citealt{Lee+16}, see the review by \citealt{Krumholz+19}). We therefore adopt $\epsilon_\mathrm{ff} = 0.01$ for this work. We emphasise that this relatively low star formation efficiency is required in order to achieve one of the goals of our project, which is to compare to dense gas tracers (e.g. NH$_{3}$) in the CMZ. A number of previous authors have adopted $\epsilon_\mathrm{ff} = 1$ \citep[e.g.][]{Torrey+17}, which is computationally-convenient because it eliminates the need to follow dense gas that evolves with small time steps. However, the price of this choice is that the dense gas obviously cannot then be studied.

\autoref{rhoSF} applies to bound, molecular gas. To apply this condition in our simulation, we allow a gas particle to be converted into a star particle if it meets all the following criteria:
\begin{itemize}
\item \textit{Self-gravitating gas.} We require gas to be locally self-gravitating. For each gas particle, we calculate the virial parameter, $\alpha$, defined as 
\begin{equation}
\alpha = \dfrac{\| \nabla \otimes {\bf{v}}_g \|^2 + (c_\mathrm{s}/h_\mathrm{g})^2}{8 \pi G \rho_\mathrm{g}}\,,
\end{equation}
with ${\bf{v}}_g$ the local gas velocity, $c_\mathrm{s}$ the adiabatic sound speed and $h_\mathrm{g}\equiv(M_\mathrm{g}/\rho_\mathrm{g})^{-1/3}$ the spatial resolution scale (average inter-particle separation around the gas particle). $\| \nabla \otimes {\bf{v}} \|^2$ is defined as 
\begin{equation}
\| \nabla \otimes {\bf{v}}_g \|^2  = \sum_{i,j=1}^3\left(\dfrac{\partial v_i}{\partial x_j}\right)^2\,.
\end{equation}
Based on this definition, $\| \nabla \otimes {\bf{v}}_g \| h_\mathrm{g}$ and $c_\mathrm{s}$ represent the local kinetic and thermal velocity dispersion, respectively.
We allow star formation only for gas particles with $\alpha <1$. This condition ensures that the gravitational energy is larger than the thermal plus kinetic energy within the resolution scale. 
\item \textit{Dense gas.} Stars are allowed to form in over-dense regions with $n_\mathrm{g} > n_\mathrm{th}$, where $n_\mathrm{g}$ is the local gas volume density and $n_\mathrm{th}$ is the star formation threshold volume density. We adopt $n_\mathrm{th} = 10^3$~cm$^{-3}$, consistent with the mean volume density in the CMZ region. 
\item \textit{Self-shielded gas.} For each gas particle, we evaluate the fraction of gas able to self-shield and cool efficiently depending on local gas column density and metallicity using the prescription by \citet{Krumholz&Gnedin11}. We allow star formation only if the fraction of self-shielded gas is larger than zero.
\item \textit{Low temperature gas.} We allow star formation only below a minimum temperature, $T<10^4$~K, where $T$ is the gas temperature. This condition ensures that a gas particle can not be converted into stars as soon as it is photoionized. 
\end{itemize}
When the above criteria are satisfied, we calculate the probability, $P$, for a gas particle to be turned into a star particle,
\begin{equation}
P = {\dot {\rho}_\mathrm{SF}} \dfrac{\Delta t}{M_\mathrm{g}}\,,
\end{equation}
where $\Delta t $ is the hydrodynamical time step and $M_\mathrm{g}$ the gas particle mass. We generate an uniform random number $N \in [0,1)$ and when $P>N$ a gas particle is converted into a star particle with same mass and dynamical properties. The star particle interacts with the gas gravitationally and via feedback (see \autoref{Cooling and Feedback}), but exerts zero pressure and feels no pressure forces.

Our resolution is high enough that the expected number of supernovae per star particle is only $\sim 1$, and the expected number of stars with significant ionising luminosity is $\ll 1$. Consequently, we cannot rely on an IMF-integrated treatment of stellar feedback. Instead, each newly-generated star particle represents an individual stellar population stochastically drawn, star by star, from a Chabrier initial mass function \citep{Chabrier05} through the stellar population synthesis code \textsc{SLUG} \citep{daSilva+12, Krumholz+15b}. \textsc{SLUG} evolves each star using the Padova stellar tracks \citep{Bressan+12}, and calculates their mass- and age-dependent ionising luminosities using the ``starburst99'' spectral synthesis method of \citet{Leitherer+99}. Stars in the initial mass range determined by \citet{Sukhbold+16} produce type II supernovae when they reach the ends of their lives.

\subsubsection{Cooling and feedback}
\label{Cooling and Feedback}

Radiative cooling is provided by the astrophysical chemistry and cooling package \textsc{GRACKLE} \citep{Smith+17}, run in equilibrium mode. GRACKLE provides cooling rates for both primordial species and metals via look-up tables calculated through the \textsc{CLOUDY} spectral synthesis code \citep{Ferland+13} as a function of
temperature and metallicity under the assumption of collisional ionization equilibrium. Gas is allowed to cool down to 10~K.
In addition to radiative cooling, we account for the presence of heating due to electrons released from dust grains by the photoelectric effect. This is implemented as a constant heating rate of $8.5 \times 10^{-26}$~erg~s$^{-1}$ per hydrogen atom uniformly throughout the simulation box \citep{Tasker&Bryan08}. This rate is consistent with the expected heating rate for the inner regions of the Galaxy \citep[$R\sim 3$~kpc,][]{Wolfire+03}. The heating rate in the CMZ region at $\sim 100$ pc is poorly constrained, but cannot be much higher than this: \citet{Ginsburg+16} used temperature-sensitive formaldehyde transitions from this region to set an upper limit on the primary cosmic ray ionisation rate $\zeta < 10^{-14}$ s$^{-1}$, which for a net heating of $\approx 15$ eV per primary ionisation \citep{Glassgold12a}, implies an upper limit on the volumetric heating rate $\Gamma \lesssim 2.5\times 10^{-25}$~erg~s$^{-1}$ per hydrogen atom, less than a factor of three from our adopted value.

In order to prevent any gas fragmentation for which the Jeans length is lower than the spatial resolution scale, we replace the thermal pressure with an artificial pressure floor, $P_\mathrm{floor} = N_\mathrm{J}^2 G \rho_\mathrm{g}^2 \,\mathrm{max}(h_\mathrm{g},s)^2/\gamma$, where $N_\mathrm{J} = 4$ is the number of elements we want to resolve. However, we do not replace the internal energy to guarantee that the gas temperature can be evolved self-consistently.
In \aref{appendix}, we conduct a study to verify how the presence of an artificial pressure affects our results.

Massive stars provide feedback to the surrounding medium through photo-ionization and supernova explosions. At each hydrodynamical time-step, \textsc{SLUG} reports the instantaneous ionizing luminosity, $\dot{N}_\mathrm{ion}$, the number of individual Type~II supernovae that occur, $N_\mathrm{SNe}$, and the mass of the ejecta, $M_\mathrm{ej}$, for each star particle.  We do not enable \textsc{SLUG} to calculate the injection rates of metal species, and thus the metallicity of the ejecta is simply $Z_\mathrm{star} M_\mathrm{ej}$, where $Z_\mathrm{star}$ is the star metallicity. Each supernova releases an energy of $10^{51}$~erg, thus we calculate the energy of the ejecta associated to each star particle as $E_\mathrm{ej} = N_\mathrm{SNe} \times 10^{51}$~erg, and the momentum as $p_\mathrm{ej} = (2 E_\mathrm{ej} M_\mathrm{ej})^{1/2}$.

\begin{figure*}
\includegraphics[width=\textwidth]{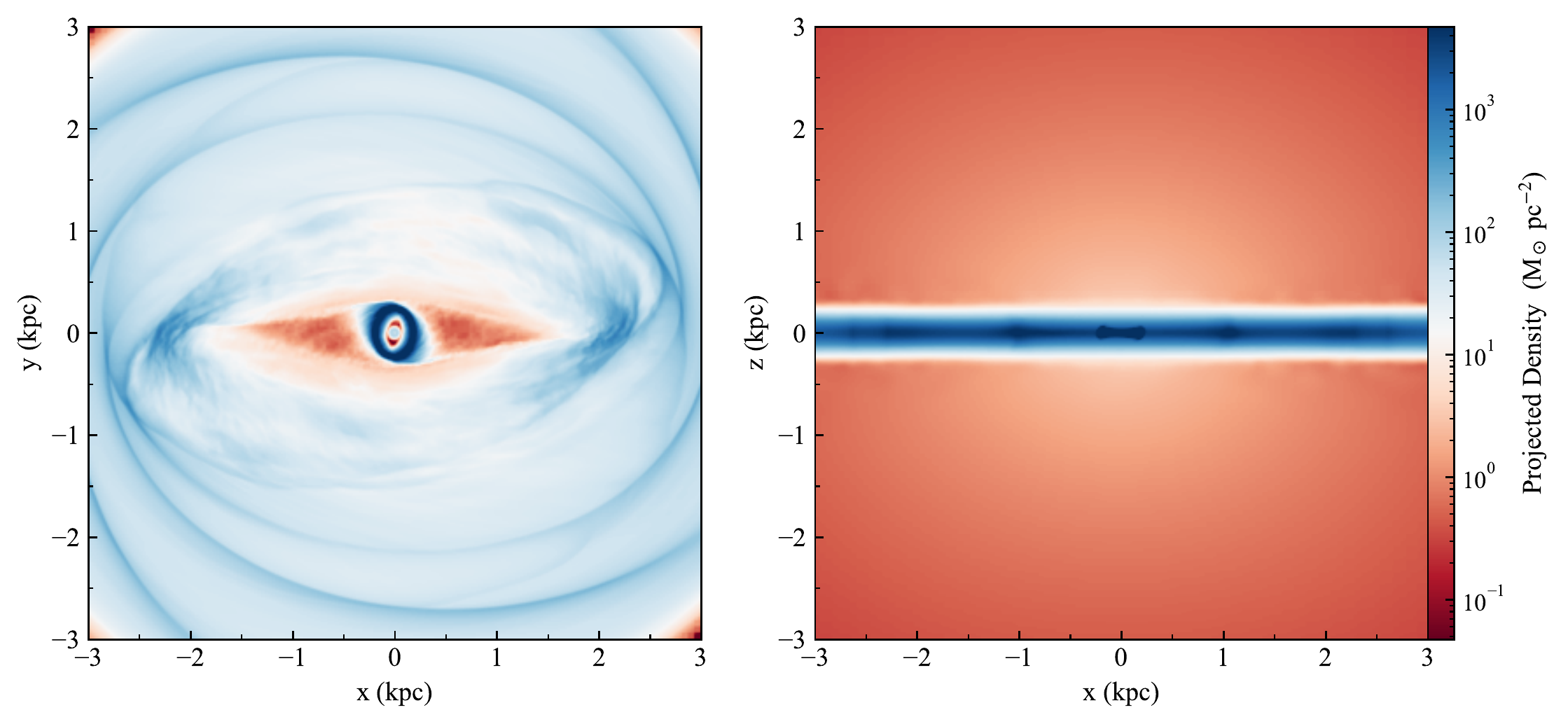}
\caption{Initial conditions of the second part of the simulation, when self-gravity, star formation, feedback and radiative cooling down to 10~K are turned on. The \textit{left panel} and \textit{right panel} show the face-on and edge-on projection of the gas density distribution in the central 3~kpc region, respectively.}
\label{IniConds}
\end{figure*}

Supernova feedback is implemented following the prescription of \citet{Hopkins+18b, Hopkins+18a}. Mass, momentum and energy are injected into the neighbouring gas particles, i.e. gas particles located within the star kernel radius or containing the star within their own kernel radius, in a fully-conservative and isotropic way. Each neighbouring gas particle gains a fraction of mass, momentum and total energy proportional to the solid angle centred on the star and subtended by the effective face between the gas particle itself and the star. Moreover, the algorithm accounts for unresolved energy-conserving phase. During this phase, the blastwave expands converting energy into momentum, until it reaches some terminal radius where the residual thermal energy has been lost and the blastwave becomes a momentum-conserving thin shell. The total amount of momentum deposited into the ambient medium is $p_t \sim 4 \times 10^5\, E_\mathrm{51}\,\mo\kms$, with $ E_\mathrm{51} = E_\mathrm{ej}/10^{51}$~erg \citep[see e.g.][]{Kim&Ostriker15, Walch&Naab15, Gentry+19}. We impose $p_\mathrm{t}$ as upper limit for the injected momentum.

Photoionization heating is also implemented following the prescription of \citet{Hopkins+18b}. The algorithm sorts all the gas particles near the star by increasing distance and, if not yet ionized (temperature below $10^4$~K), it calculates the ionization rate needed to ionize it, $\Delta \dot{N}_\mathrm{ion} (\rho_\mathrm{g})$. If $\Delta \dot{N}_\mathrm{ion}\leq \dot{N}_\mathrm{ion}$ or $\Delta \dot{N}_\mathrm{ion}/\dot{N}_\mathrm{ion} > x$ with $x \in [0,1)$ a uniform random number, the gas temperature is set to $10^4$~K and the ionizing luminosity is replaced by $\dot{N}_\mathrm{ion}\longmapsto \dot{N}_\mathrm{ion} - \Delta \dot{N}_\mathrm{ion}$. We then proceed to the next particle and repeat the process until $\dot{N}_\mathrm{ion} \sim 0$. Any photoionized gas particle is not allowed to cool during the entire star time-step.

We do not account for the presence of a super massive black hole and its feedback in the Galactic centre, since at this stage we are interested in investigating feedback processes associated to star formation activity only.

\subsection{Initial conditions}
\label{Initial conditions}

The initial conditions of the simulation include gas particles only. We initially generate $2\times10^6$ gas particles with mass $2\times10^3~\mo$; in \aref{append} we conduct convergence studies to verify that our results are robust against changes in mass resolution. The gas particles are sampled in order to reproduce the mass distribution in the Galactic disc given by \citet{Binney&Tremaine08},
\begin{equation}
M_\mathrm{g} (R, z) = 2 \pi \int_{-z}^z \int_0^R R\,  \rho_\mathrm{g} (R, z)\, dR \,dz\;,
\end{equation}
with
\begin{equation}
\rho_\mathrm{g} (R, z) = \dfrac{\Sigma_\mathrm{g}}{2z_\mathrm{g}} \rm{exp} \left(-\dfrac{R}{R_\mathrm{g}} - \dfrac{\vert z\vert}{z_\mathrm{g}}\right)\;,
\label{MWdens}
\end{equation}
where $\Sigma_\mathrm{g} = 131.8\, \mo \, \rm{pc}^{-2}$ is the central surface gas density, $z_\mathrm{g} = 80$~pc is the gas scale height and $R_\mathrm{g} = 4.5$~kpc is the gas scale radius. Gas particles are distributed inside a cylindrical slab with radius 4.5~kpc and half-height 1~kpc. The total gas mass in this region is $\sim 4 \times10^9\,\mo$. The velocity field is initialized so that gas particles orbit in the Galactic plane with circular velocity equal to $V_\mathrm{c}(R)$ (black line in \autoref{Rotcur}) and vertical velocity equal to zero. We assume that gas has a uniform temperature of $10^4$~K and solar metallicity, regardless of Galactocentric radius. The reader should note that in \autoref{MWdens} we have omitted the term  exp($-R_\mathrm{m}/R$), present in \citet{Binney&Tremaine08}. This term accounts for the presence of a hole of radius $R_\mathrm{m} = 4$~kpc, due to the presence of the Galactic bar that sweeps out gas from the bar-dominated region. However, we prefer to start the simulation with gas in radial equilibrium in the axisymmetric gravitational potential of the Milky Way and let it slowly adjust to the non-axisymmetric part of the potential. 

Beyond the gas distribution in the Galactic disc, we also set up the initial conditions for the rarefied and hot halo surrounding the Galaxy. We generate $3\times10^4$ gas particles, with the same mass as the disc gas particles, $2\times10^3~\mo$, and we sample them in order to be in hydrostatic equilibrium in the axisymmetric part of the gravitational potential. We set the halo temperature and metallicity to $2\times10^6$~K and 0.1~Z$_\odot$, respectively, consistent with observations \citep{Sembach+03, Miller&Bregman15} and we set the central volume density in order to have pressure equilibrium with the disc. The total halo gas mass within 4.5 kpc from the Galactic centre is $\sim 6 \times 10^7\,\mo$. 

We run the simulation for 300~Myr to allow the system to reach a steady state equilibrium in the non-axisymmetric potential. In this first part of the simulation self-gravity, star formation and stellar feedback are turned off, so that gas evolves in the presence of pure hydrodynamics and external gravity. Moreover, gas is allowed to cool down to temperature of $10^4$~K only in the galactic disc, and radiative cooling is disabled entirely in the halo; we turn off cooling because, in the absence of a heating source or stellar feedback, dense gas in the innermost part of the halo cools down, thus breaking the condition of hydrostatic equilibrium. Reflecting boundaries are located at $x,y,z = \pm 6$~kpc.

Once a steady state equilibrium is reached, we combine the disc and halo gas distributions and turn on self-gravity, star formation, stellar feedback and full radiative cooling. \autoref{IniConds} shows the initial conditions of this second part of the simulation in the central 3~kpc region. The left panel shows the face-on gas density projection. The gas dynamics in the central region of our Galaxy has been briefly explained in  \autoref{Introduction}. Due to the presence of the rotating Galactic bar, gas flows from the outer parts of the disc towards the centre following $x_1$ orbits. At $\sim2$~kpc from the centre along the bar major axis gas is shocked to the CMZ and settles into a high-density (volume density $\sim 10^{4}$~cm$^{-3}$) ring at $\sim 250$~pc from the Galactic centre. The transition happens through the so-called dust lanes (colour-coded in white), located in the low-gas density region between the CMZ and the innermost $x_1$ orbit. 
We point out that the size of the ring formed in our simulation is a factor of 2 larger than the observed size of the CMZ ring (radius $\sim 100-150$~pc, see \autoref{Introduction}). The gravitational potential used in our simulation (\autoref{Potential}) was built to be consistent with large scale properties of the Milky Way, but it was not fine-tuned to reproduce the size of the CMZ ring \citep[see also][]{Sormani+18}. Only recently \citet{Sormani+19b} have modified - compatibly with observational constraints - the bar parameters in the gravitational potential to match the observed ring size.

Contrary to the predictions of \citet{Krumholz+15a} and \citet{Krumholz+17}, the location of the dense ring is not at the shear minimum \citep[see also][who first proposed this theory]{Lesch+90}. One possible explanation for this discrepancy is that gas does build up a ring where the shear reaches a minimum and, if it is able to cool, it begins to fragment into clouds. In the absence of feedback, all the mass goes into these dense clouds, which are essentially collisionless with respect to one another, and thus cannot effectively transport angular momentum; they therefore remain where they form. On the other hand, if gas is not entirely collected into effectively-collisionless clouds, either due to an artificial cooling floor (as in our first phase) or due to star formation feedback (as we will see below), then angular momentum transport is possible. Consequently, the majority of the material moves somewhat inwards from the initial location of ring formation, while a small amount moves outward. Alternatively, the mismatch between shear minimum and ring location could provide evidence in favour of theories predicting that the ring size is controlled only by bar properties (e.g. bar strength and patter speed) and effective sound speed of the gas \citep[e.g.][]{Kim+12, Sormani+15, Sormani+18}, independent of the shear. In this model, at fixed bar parameters, the ring size decreases with increasing the gas sound speed. This might explain why, in the simulation of \citet{Sormani+17} using the same rotation curve adopted here, the ring size  ($\approx 400$~pc) is larger than in our simulation -- \citet{Sormani+17} do not include feedback, and thus the majority of the gas rapidly cools down to 10~K.

The right panel of \autoref{IniConds} shows the edge-on gas density projection, where the presence of the Galactic halo is visible (colour-coded in red). The halo is almost isothermal, with temperature around  $2\times10^6$~K. Its central volume density is $\sim 0.3$~cm$^{-3}$, while its volume density at 4.5 kpc above the plane is $\sim 10^{-3}$~cm$^{-3}$. The dense ring is visible in projection.

\begin{figure*}
\includegraphics[width=\textwidth]{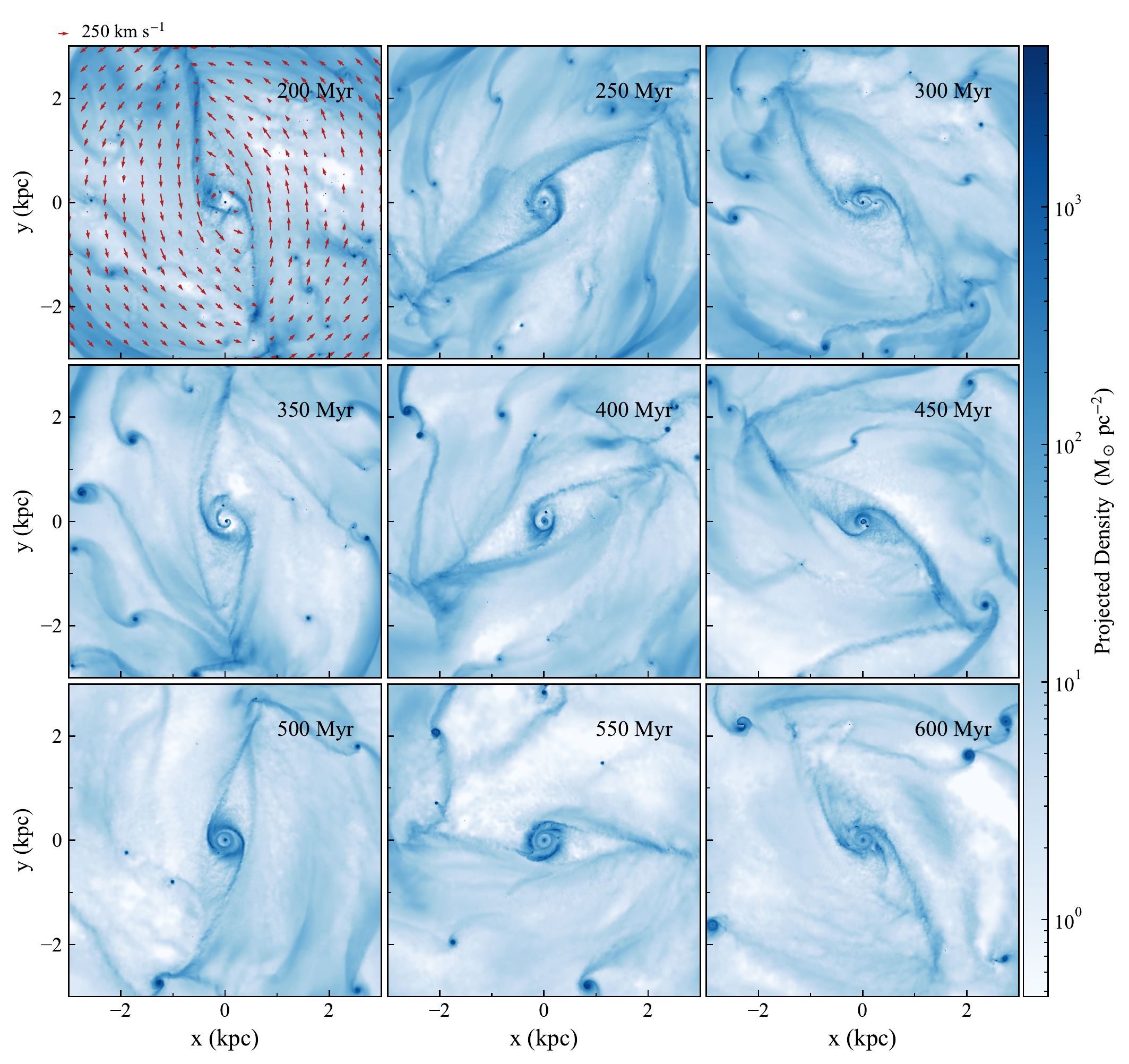}
\caption{Temporal evolution of the face-on gas density projection in the central 3~kpc region of the Galaxy. The time at which the snapshots have been taken is indicated in the top right corner of each panel. In the top left panel the velocity field at $z=0$ overlaps the density projection. The velocity field is shown with vectors, whose length indicates the velocity magnitude.}
\label{FullDisc}
\end{figure*}

\begin{figure*}
\includegraphics[width=\textwidth]{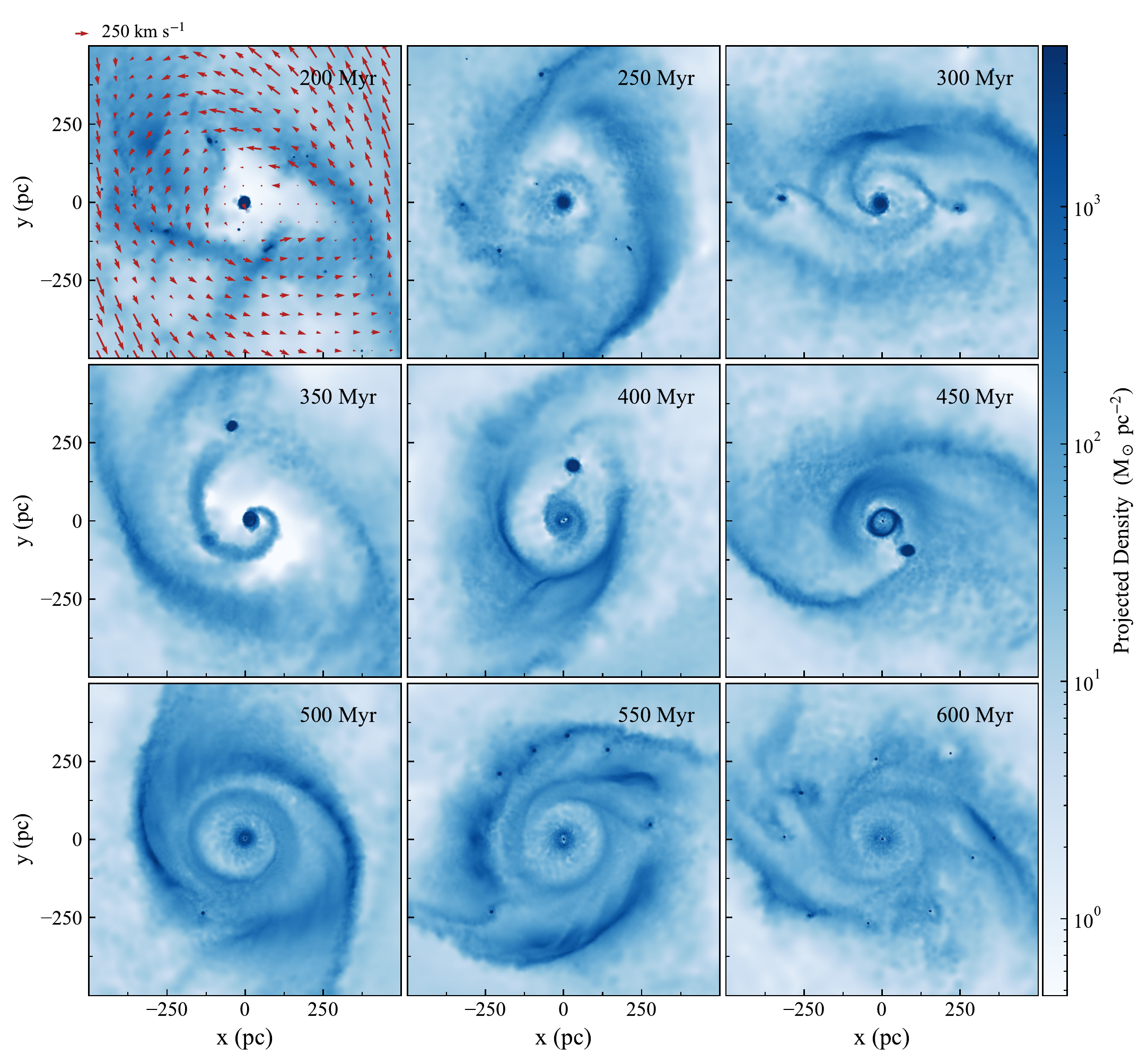}
\caption{Same as \autoref{FullDisc}, but zooming in on the central $500$~pc of the Galaxy.}
\label{CMZ}
\end{figure*}

\begin{figure}
\includegraphics[width=0.49\textwidth]{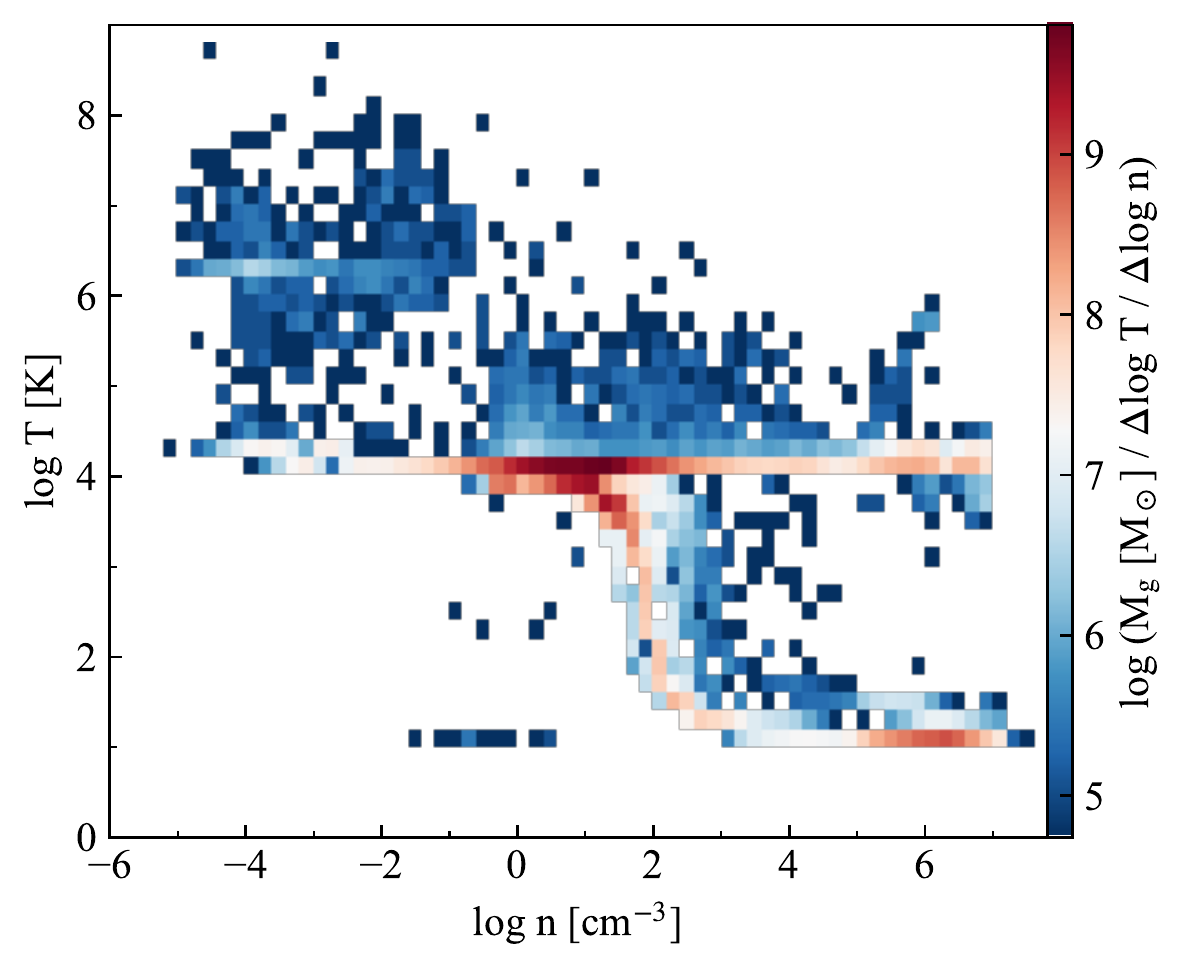}
\caption{Temperature-density phase diagram of the entire Galaxy at $t=300$~Myr.  The diagram is computed as a two-dimensional histogram showing the mass of gas particles within each logarithmic bin, normalised by the bin area. }
\label{PhasePlot}
\end{figure}

\section{Results}
\label{Results}

\begin{figure*}
\includegraphics[width=\textwidth]{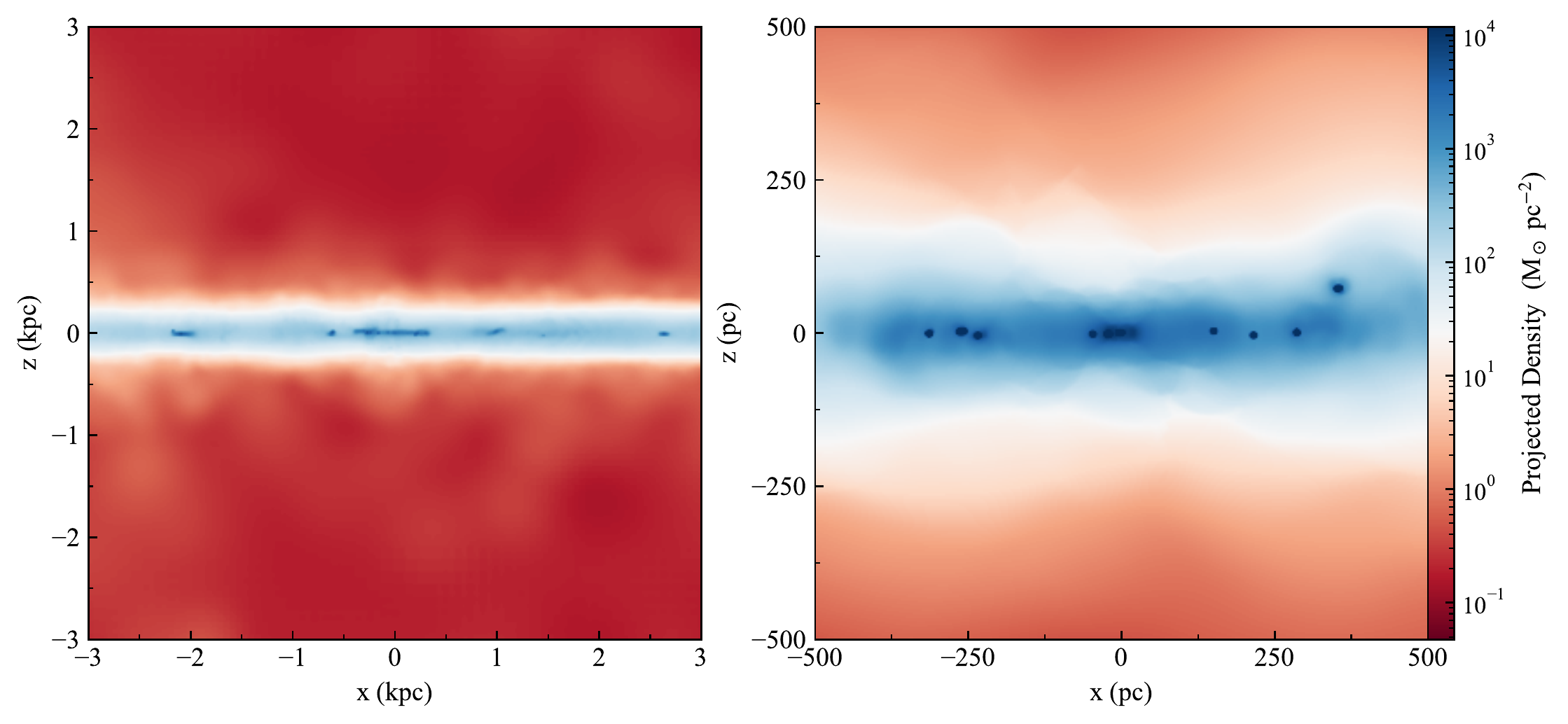}
\caption{Edge-on density projection in the central 3~kpc \textit{(left panel)} region and 500~pc \textit{(right panel)} region of the Galaxy. The snapshots have been taken at $t=600$~Myr.}
\label{CMZ_edgeon}
\end{figure*}

In the first part of the simulation the gas distribution within the CMZ ring is very smooth due to the absence of turbulent motions. Thus, as soon as we turn on self-gravity, most of this gas collapses in a few hundred kyrs, quickly encountering the conditions required for forming stars (see \autoref{SF}). This generates a burst of star formation during the first few Myrs of the simulation, after which stellar feedback starts to be effective and slowly pushes the system back towards equilibrium. In \autoref{SFR}, we quantitatively analyse the star formation and the gas mass evolution as a function of time and we conclude that a quasi steady-state equilibrium is reached at 200 Myr after the onset of star formation. We therefore focus the bulk of our analysis on the evolution subsequent to this point.

\subsection{The CMZ morphology}
\label{Morphology}

\autoref{FullDisc} and \autoref{CMZ} show face-on gas density projections in the inner 3~kpc and 500~pc regions respectively. Density snapshots are taken at different times starting from 200 Myr after the onset of star formation, i.e. when a quasi steady-state equilibrium is reached. The velocity field at $z=0$ overlaps the gas density distribution in the snapshot at $t=200$~Myr.
The zeroth-order gas dynamics described in \autoref{Introduction} and \autoref{Initial conditions} is still valid. Gas flows from the outer parts of the Galactic disc towards the CMZ through the dust lanes. The latter extend from the CMZ up to $\sim 2-3$~kpc  along the bar major semi-axis and are clearly visible at any time, suggesting that the gas flow is continuous across time. However, in contrast with \autoref{IniConds}, where the disc exhibits a gas distribution point-symmetric with respect to the Galactic centre, here the gas distribution is broken up into molecular clouds and connecting filaments, whose centre of mass nearly follows $x_1$ and $x_2$ orbits depending on the disc regions. 

The formation of molecular clouds at these larger radii is a result of the combined effects of thermal instability and the Galactic bar. To illustrate this, in \autoref{PhasePlot} we show the temperature-density phase diagram of the Galaxy at $t=300$ Myr. We see that most of the gas in the simulation lies at temperature $T\sim10^4$~K and density $n \sim 1-10$~cm$^{-3}$. This represents stable warm neutral medium (WNM; note that our equilibrium WNM temperature and density are somewhat higher than is typical in the Solar neighbourhood due to the somewhat higher photoelectric heating rate we adopt for the Galactic Centre). Almost all of the diffuse medium at Galactocentric radii of $\sim 2-3$ kpc is in this phase, and examination of \autoref{IniConds} or \autoref{FullDisc} shows that the typical surface density everywhere except inside a molecular cloud or at the tip of the bar is $\Sigma\sim 30\,\mo$. Consequently, the Toomre $Q$ parameter $Q \approx \sqrt{2} \Omega c_s / \pi G \Sigma$ is of order a few, so the gas is stable. Cloud formation occurs for two reasons. First, gas accumulates at shocks, particularly at the tip of the bar, causing the surface density to rise by a factor of $\sim 10$. Second, compression of gas in these regions drives the volume density up to $n \gtrsim 100$~cm$^{-3}$, at which point forbidden line cooling overwhelms photoelecetric heating and the gas transitions to cold neutral medium (CNM) at $T\lesssim100$~K (\autoref{PhasePlot}) \footnote{The negligible amount of gas at $T\sim10$~K and $n \sim 0.01-1$~cm$^{-3}$ is not CNM, but material undergoing highly-supersonic rarefaction. Its location in such region of the phase diagram is transient, since this gas will be quickly heat up to $10^4$~K under the effect of photoelectric heating.}. Thus shocks both lower $c_s$ and raise $\Sigma$, driving the gas to $Q \ll 1$ so that it fragments into clouds.\footnote{Note that a corollary of this analysis is that we can strongly rule out the possibility that the fragmentation is driven by numerical instabilities. At the $\sim 10^4$ K temperatures that prevail in the diffuse medium at $2-3$ kpc, the Toomre mass is $\sim 10^7\,\mo$, a factor of almost $10^4$ larger than our particle mass. Thus we resolve the scales of gravitational instability extremely well until the gas undergoes the WNM-CNM phase transition due to compression.} The presence of massive clouds co-rotating with the disc causes the formation of swing-amplified shearing flows (visible as spiral wavelets extending from the dust lanes to the outer disc) which wrap and feed the clouds enhancing their level of rotation (up to 100 km s$^{-1}$ in some cases). 
However, we would like to note that the internal structure of the molecular clouds is not properly resolved in our simulation: in the absence of softened gravity (we remind that gravity is always softened on scales lower than 1~pc) and pressure floor the clouds would presumably fragment into smaller structures with a different internal dynamics. Thus, the high-level of rotation produced for some clouds in the simulation is most likely unreliable.

\autoref{CMZ} focuses on the gas evolution in the CMZ region. The morphology of the CMZ is much more complicated than the one displayed in \autoref{IniConds}.
At different times, the CMZ presents either a ring-like (see e.g. snapshots at $t=200$~Myr) or a spiral-like (see e.g. snapshots at $t=350$~Myr) morphology or both of them (see e.g. snapshots at $t=500$~Myr). In all cases, most of the gas is located in streams, whose mass-weighted Galactocentric radius varies between 200 to 300~pc (see also \autoref{Gas&SFRRing}). 
As discussed in \autoref{Initial conditions}, the ring size can depend on the gas sound speed, decreasing with increasing the sound speed. In the second part of the simulation, the gas is allowed to cool down to 10~K, but we see no evidence that the location of the gas ring is altered by the onset of cooling. This may be because a significant mass of gas in the nuclear ring is photoionized and heated up to at $T \sim 10^4$~K (see \autoref{PhasePlot}), producing an effective sound speed similar to that produced by the cooling floor of $10^4$~K during the initial, setup phase of the simulation. The gas pressure support is also provided by supernova-feedback that injects small-scale turbulence into the ISM. Besides shrinking the gas towards the Galactic centre, another effect of the high gas pressure, both thermal and turbulent, should be to favour the development of spiral shocks rather than ring-like structures \citep[see][]{Sormani+18}.

Obviously, due to the non-uniform distribution of star formation, the above phenomena cannot be point-symmetric with respect the Galactic centre but depend instead on local conditions of stellar feedback, radiative cooling and gravity. The result is that the gas distribution in the CMZ is highly asymmetric, in agreement with observational findings (see \autoref{Introduction}). Asymmetries within the CMZ ring were also found by \citet{Sormani+17}, even though their simulations were run in the absence of self-gravity and stellar feedback. In such case, asymmetries develop due to the combination of thermal instability and the so-called ``wiggle instability''.

In \autoref{CMZ}, we can also note that the diffuse gas streams contains star-forming molecular clouds, whose number, mass and size change with time. At different times, the CMZ can be comprised of about ten clouds with masses around $10^{5-6} \, \mo$ (see e.g. snapshots at $t=550$~Myr), a single, dominant very large cloud with mass $\lesssim 10^{7} \, \mo$ (see e.g. snapshots at $t=400$~Myr), or a predominantly diffuse medium with no clouds at all (see e.g. snapshots at $t=500$~Myr).

\begin{figure*}
\includegraphics[width=\textwidth]{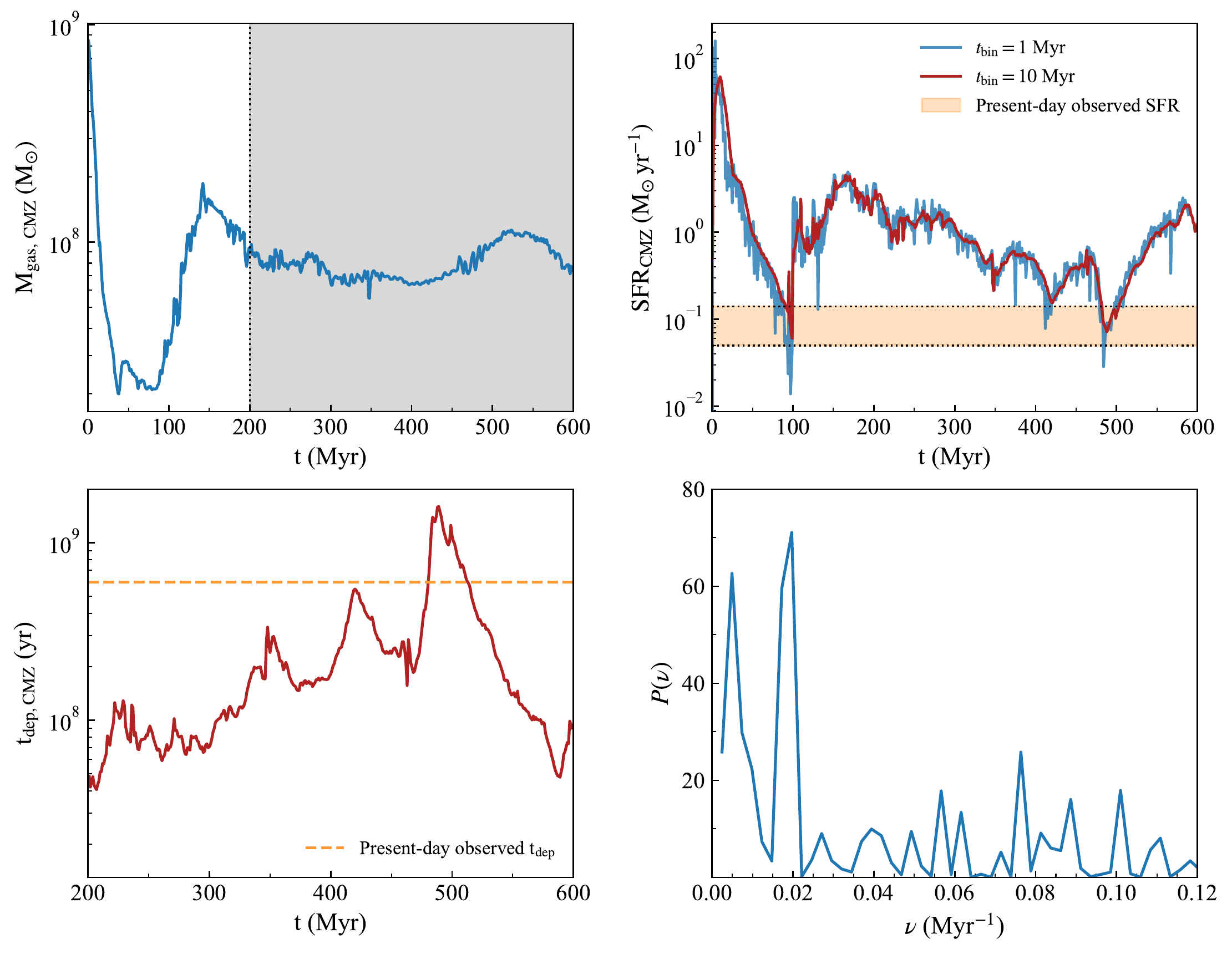}
\caption{Time evolution of gas mass (\textit{top left panel}),  SFR (\textit{top right panel}) and depletion time (\textit{bottom left panel}) in the CMZ and power spectrum of SFR$(t)$ (\textit{bottom right panel}). The grey region in the top left panel detonates the range of time over which the mass in the CMZ is roughly constant and a quasi-steady state equilibrium has been reached. In the top right panel, the blue and red lines indicate the SFR averaged over the past 1~Myr and 10~Myr respectively, while the orange bar indicates the approximate present-day observed SFR. In the bottom left panel, the red lines detonates the gas depletion time across the simulation, while the orange dotted line indicates the present-day CMZ depletion time calculated by \citet{Kruijssen+14}. The power spectrum in the bottom right panel is evaluated over the time window between 200 and 600~Myr.}
\label{SFR&GasMass}
\end{figure*}

The left and right panel of \autoref{CMZ_edgeon} show the edge-on density projections (integrated over the \textit{y-}axis) in the inner 3~kpc and 500~pc regions of the Galaxy, respectively. We show these views only at a single time because, viewed edge-on, the morphology does not vary dramatically over time. Cold and warm neutral gas ($T\lesssim10^4$~K) corresponds to projection densities $\gtrsim 10^2\, \mo$~pc$^{-2}$ in the CMZ region ($\lesssim 500$~pc from the Galactic centre, as shown in the right panel) and $\gtrsim 10 \,\mo$~pc$^{-2}$ in the region from $\approx 0.5-3$~kpc shown in the left panel. Therefore, the thickness of the neutral disc is much lower in the CMZ region, $h\sim 80$~pc, than at larger Galactocentric radii, $h\sim 300$~pc, where the gravitational potential is weaker. Volume densities high enough that we would expect the gas to be predominantly molecular correspond to projected densities $\gtrsim 10^3\, \mo$~pc$^{-2}$ and the disc scale height of this material is $\sim10-20$~pc only. We carry out a much more detailed investigation of the partition of the CMZ into different chemical phases, and the observable properties of the molecular and atomic line emission, in later paper in this series.

\subsection{Gas mass and star formation evolution}
\label{SFR}

We next analyse how the gas mass and the SFR in the CMZ evolve with time. We evaluate these two quantities within a cylinder centred at the Galactic centre, with radius 500~pc and half-height, $h/2$, 200~pc, corresponding to roughly twice the scale height of the gaseous disc ($h\sim 80$~pc, see \autoref{Morphology} and \autoref{CMZ_edgeon}) in the CMZ region. The SFR is computed as a running average over a time window $\Delta t_\mathrm{bin}$ as follows,
\begin{equation}
\mathrm{SFR}(t) = \sum \dfrac{M_\mathrm{star} (t_\mathrm{age} (t) < \Delta t_\mathrm{bin})}{\Delta t_\mathrm{bin}}\;,
\label{eq:SFR}
\end{equation}
where $M_\mathrm{star}$ is the mass of an individual star particle and $t_\mathrm{age}(t)$ is the stellar age at a given time. We choose two different values of $\Delta t_\mathrm{bin}$, 1~Myr - which is the time step of the simulation outputs - and 10~Myr. The latter has been chosen to permit a fairer comparison with the observations, since observational SFR indicators, as H$\rm{\alpha}$ and FUV emission, trace stars formed over the past $\sim 10$~Myr. 

The top left panel of \autoref{SFR&GasMass} shows the time evolution of gas mass. The amount of mass decreases by almost 2 orders of magnitude in the first $\sim 50$~Myr. As soon as we turn on self-gravity, all the hydrodynamical quantities are very smooth within the CMZ ring. The absence of turbulent motions that might hinder self-gravity causes a sudden collapse of the entire CMZ ring and a strong burst of star formation. Stellar feedback associated to this burst is very effective and depletes the CMZ of gas in a few ten of Myrs. After almost $100$~Myr from the onset of star formation, gas starts again to flow from the outer regions of the disc towards the centre. The amount of gas mass in the CMZ increases with time until becoming roughly constant after $\sim 200$~Myr. In this range of time (grey region), the gas mass varies by less than a factor of 2. The mass of gas in the CMZ lies in the range between $6\times10^7$ and $10^8 \,\mo$, in agreement with the observed values ($3-7\times10^7 \,\mo$ for molecular gas, see \autoref{Introduction} for references). Since the initial burst of star formation is due to unrealistic initial conditions, in the following analysis we calculate the gas properties after 200~Myr only.

The top right panel of \autoref{SFR&GasMass} shows the SFR as a function of time. The blue and red lines indicate the SFR averaged over the past 1~Myr and 10~Myr, respectively. In the range of time after 200 Myr, star formation activity goes through oscillatory cycles and the SFR varies between $2 \moyr$ to a few times $0.01 \moyr$. In order to evaluate whether there are characteristic variability time-scales, in the bottom right panel of \autoref{SFR&GasMass}, we calculate the power spectrum of SFR$(t)$ over the time window between 200 and 600~Myr. We find that the power spectrum peaks around temporal frequencies, $\nu$, of 0.005 and 0.02~Myr$^{-1}$, corresponding to $\Delta t \sim 200$~Myr and $\Delta t \sim 50$~Myr. Star formation cycles on these time-scales are clearly visible in \autoref{SFR&GasMass}. The SFR decreases by more than one order of magnitude in the temporal range between $\sim 250$~Myr and $\sim 480$~Myr. After that time, the SFR increases again up to values of $2 \moyr$. Within this large cycle, there are a few shorter star formation cycles of the order of $50-100$~Myr. 

The short-period cycle of $\approx 50$~Myr is most likely driven by feedback instabilities. The high densities in the CMZ causes bursts of star formation, which end when supernova feedback become fully effective, almost $40-50$~Myr after the onset of star formation. 
However, this feedback does not remove the bulk of the ISM from the CMZ, since, as shown in the upper left panel of \autoref{SFR&GasMass}, the gas mass in the CMZ is nearly constant even as the SFR varies by more than an order of magnitude. Instead, the mechanism for variation is that feedback greatly increases the gas velocity dispersion and this disrupts the dense star-forming clouds and the dense ring, consistent with the scenario proposed by \citet{Krumholz+17}.

The long-period variation ($200$~Myr) is less straightforward to explain, since we do not run the simulation for a time long enough to establish whether it is a stochastic episode only. One possible explanation is that this is due to rapid migration of a large molecular cloud ($M\lesssim10^7\,\mo$) towards the Galactic centre (see snapshots between $t=350$~Myr and $t=450$~Myr in \autoref{CMZ}). The migration time of this cloud  is $t_\mathrm{migr} = 2.1 \,Q^2 \delta^{-2} / \Omega \approx 150$~Myr \citep[][]{Dekel+09}, where $Q\approx1$ is the Toomre parameter, $\Omega\sim \,600 \kms$~kpc$^{-2}$ is the angular velocity at the ring position, $R\sim 250$~pc, and $\delta\sim 0.16$ is the ratio between the cloud mass and the gas mass within $R\sim 250$~pc. The rapid migration of the cloud is due to the high ratio between the cloud mass and the gas mass in the CMZ and the high angular velocity.\footnote{Note that the bar is probably unimportant for inducing transport this close to the Galactic Centre. Inside 250 pc, the bar does not dominate the overall potential (c.f.~\autoref{Rotcur}), and the pattern speed of the bar is an order of magnitude smaller than the orbital speed.} During its trajectory, the cloud slowly loses mass by tidal stripping \citep[see e.g.][]{Dekel+13} causing a decline in the overall star formation budget. In the simulation, the migration ends at $\sim 450$~Myr when the cloud is tidally destroyed by its encounter with the Galactic centre. If such large clouds form and then migrate regularly, that would cause periodicity on the migration timescale. However, since we only run the simulation long enough to see a few such episodes, we cannot draw more than tentative conclusions.

The orange bar in the top right panel of \autoref{SFR&GasMass} indicates the approximate present-day observed SFR ($\sim 0.04-0.1 \moyr$, see \autoref{Introduction} for references). On average, the SFR in the simulation is higher than the observed one. However, several lines of evidence indicate that the present-day CMZ is likely at a minimum of a star formation cycle (e.g. $t=400$~Myr or $t=480$~Myr in the simulation). In the bottom left panel of \autoref{SFR&GasMass}, we calculate the gas depletion time in the CMZ, $t_\mathrm{dep} = M_\mathrm{gas}/\mathrm{SFR}$, as a function of time, using the SFR averaged over a 10~Myr window. The red line shows the depletion time evolution across the simulation, while the dotted orange line indicates the present-day depletion time calculated by \citet{Kruijssen+14} using observational estimates of molecular gas and SFR surface densities. In the simulation, $t_\mathrm{dep}$ varies by more than one order of magnitude, from a few~$\times \,10^7$ to $10^9$~yr. Its time-averaged value is $\sim 2 \times 10^8$~yr, in agreement with the average depletion time found by \citet{Leroy+13} in the dense nuclear regions of some nearby disc galaxies. The present-day depletion time of the CMZ, $\sim 6 \times 10^8$~yr, is consistent with our results, although it lies at the outskirts of the range of depletion times produced during the simulation, in a region where the star formation activity goes through a quenching period.

\subsection{Gas mass and star formation distribution}
\label{Gas&SFRRing}

\begin{figure}
\includegraphics[width=0.49\textwidth]{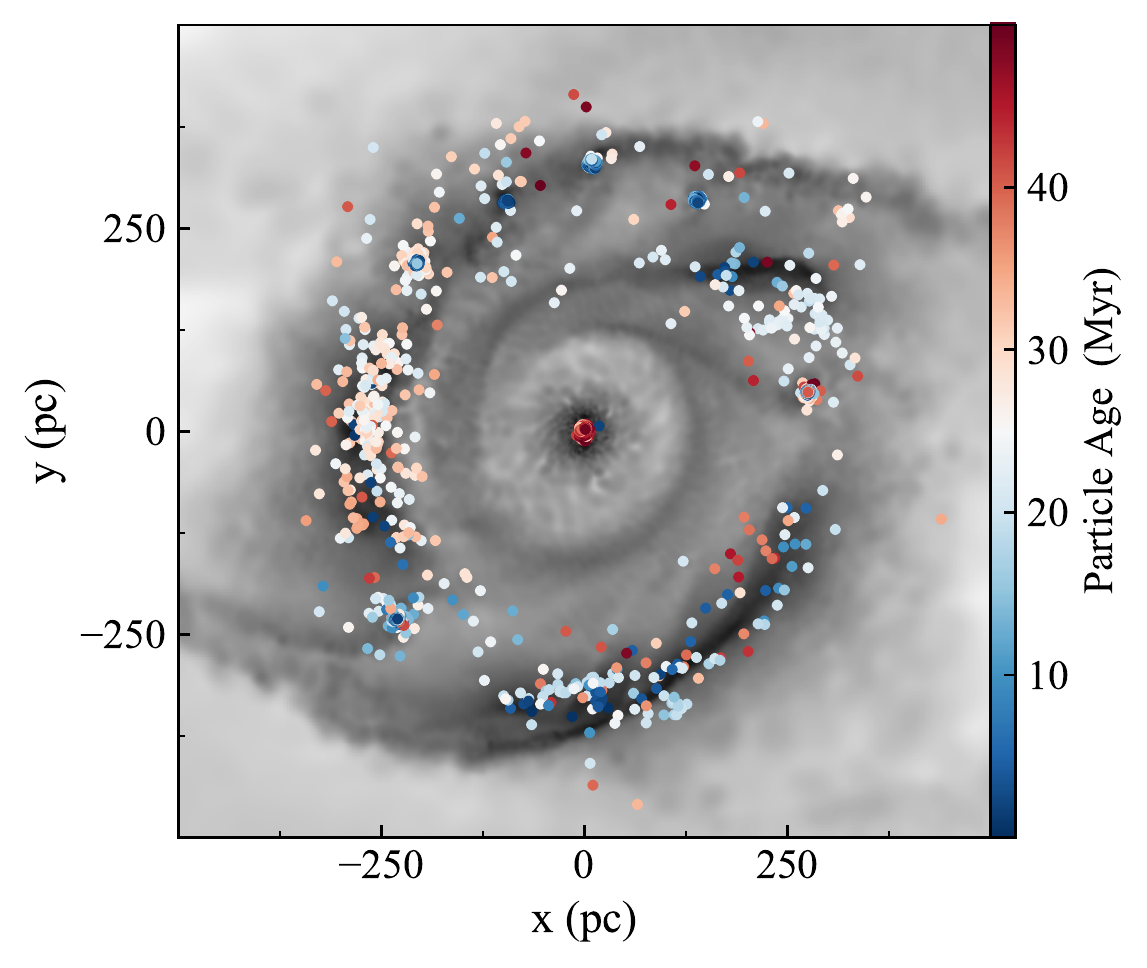}
\caption{Star particles with ages $<50$ Myr (coloured dots) plotted over the projected density distribution (greyscale map) in the central 500~pc region of the Galaxy. The colour of each star indicates its age. The snapshot has been taken at $t=550$~Myr.}
\label{Stars}
\end{figure}

\begin{figure}
\includegraphics[width=0.48\textwidth]{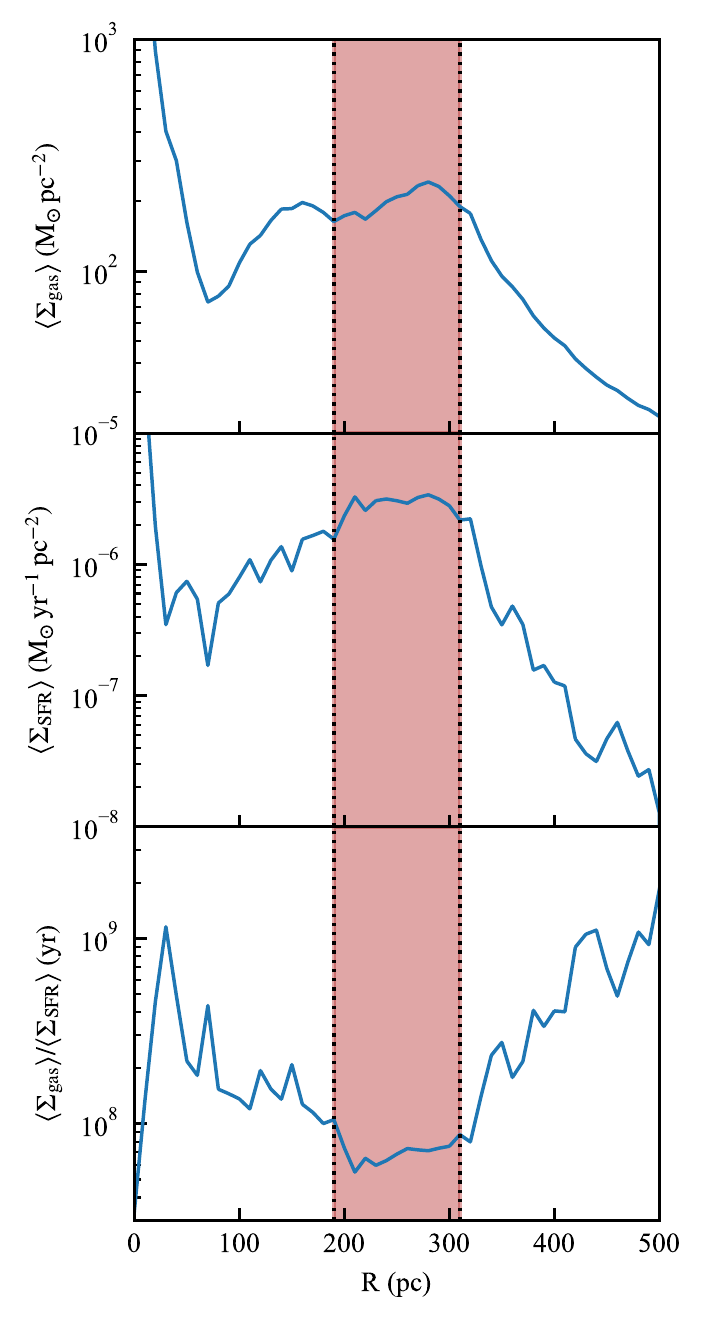}
\caption{Time-averaged gas surface density ($\langle\Sigma_\mathrm{gas}\rangle$, \textit{top panel}) and SFR surface density ($\langle\Sigma_\mathrm{SFR}\rangle$, \textit{middle panel}) and their ratio (\textit{bottom panel}) as a function of Galactocentric radius. The red bar indicates the range of values where the mass-weighted radius lies.}
\label{Gas&SFRdistribution}
\end{figure}

In addition to the time evolution of gas mass and SFR, we are interested in analysing their spatial distribution. As a first step, in \autoref{Stars} we show the distribution of star particles younger than 50 Myr over-plotted on the gas distribution in the CMZ at $t=550$~Myr. We see that young star particles are located within the CMZ ring, and are almost absent elsewhere (except for the the central few parsec region), implying that the nuclear ring is the region where star formation occurs. If we consider newly-formed stars, i.e. stars younger than 5 Myr (colour-coded in blue), we see that the bulk of them reside within clouds (see cloud location in \autoref{CMZ}), meaning that most of the stars form within these structures.

\begin{figure*}
\includegraphics[width=\textwidth]{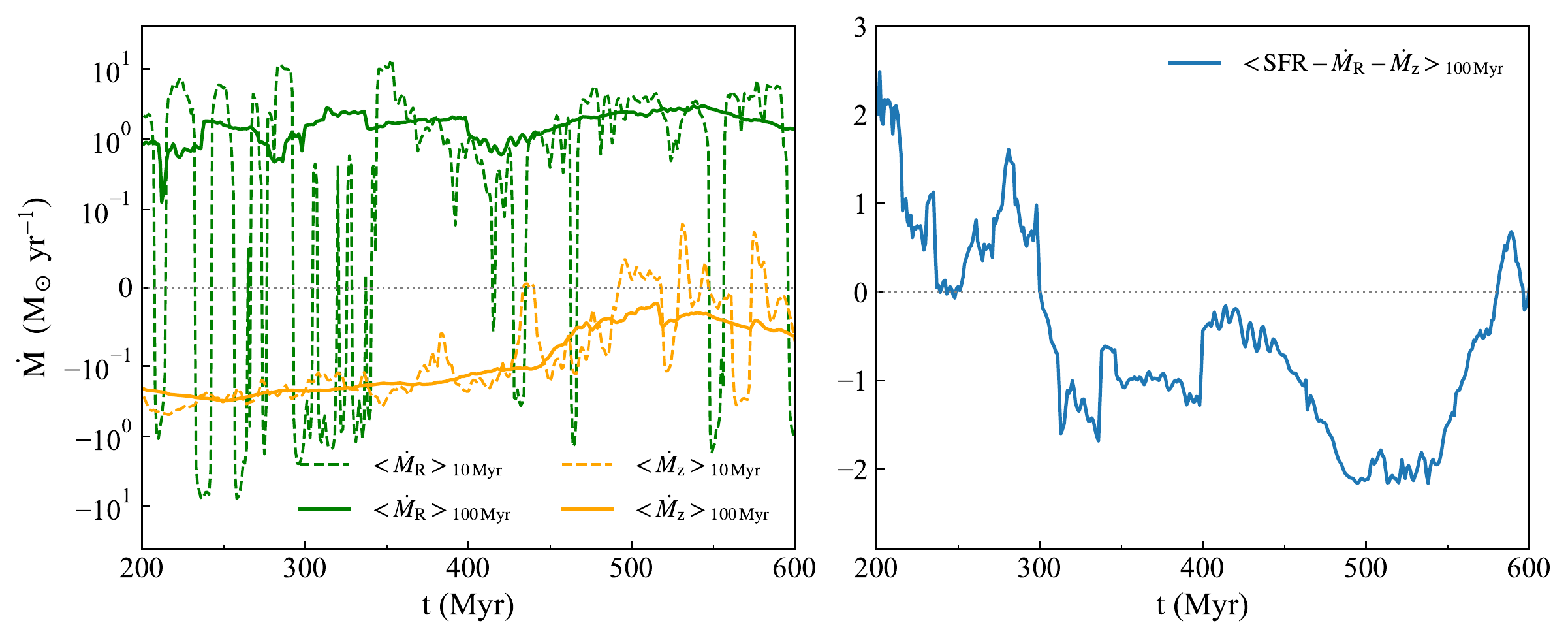}
\caption{\textit{Left panel}: time evolution of the mass flow rates (inflow mass rate - outflow mass rate) across the lateral surface ($\dot{M}_\mathrm{R}$, green lines) and the top/bottom surfaces ($\dot{M}_\mathrm{z}$, yellow lines) of the cylinder containing the CMZ ($R = 500$~pc, $h /2= 200$~pc). The mass flow rates are convolved with 1D box functions with width 10~Myr (dashed lines) and 100~Myr (solid lines). The $y$-axis is shown in symmetrical logarithmic scale, corresponding to linear scale for values from 0 to $\pm 0.1$, and logarithmic scale for values larger/lower than $\pm 0.1$. \textit{Right panel}: time evolution of the net gas mass variation within the CMZ convolved with a 1D box function with width 100~Myr.}
\label{NetFlux}
\end{figure*}

To make this analysis more quantitative, we evaluate the amount of gas and SFR within radial bins and divide these two quantities by the area of each bin, thus obtaining the radial distribution of gas surface density, $\Sigma_\mathrm{gas}$, and SFR surface density, $\Sigma_\mathrm{SFR}$.
\autoref{Gas&SFRdistribution} shows the time-averaged distribution of $\Sigma_\mathrm{gas}$ (top panel) and $\Sigma_\mathrm{SFR}$ (middle panel) as a function of Galactocentric radius. $\Sigma_\mathrm{gas}$ and $\Sigma_\mathrm{SFR}$ present a similar trend. They exhibit very high values in the Galactic centre, where the gravitational potential well of the Galaxy holds a significant fraction of gas mass (a few~$\times \,10^6\, \mo$) in a region of a few parsecs. This region is unresolved in our simulations, and since we do not include the gravitational potential of the black hole that dominates the central $\sim1$~pc, the dynamics within it are not credible in any event. However, this region also contributes little to the overall mass or star formation budget of the simulation, and therefore we will not discuss it further. Outside this unresolved region, the gas surface density then decreases by more than one order of magnitude in the first 80~pc. Beyond this region, $\Sigma_\mathrm{gas}$ and  $\Sigma_\mathrm{SFR}$ increase until reaching an almost constant value ($\Sigma_\mathrm{gas} \sim 200 \mo$~pc$^{-2}$ and  $\Sigma_\mathrm{SFR} \sim 3 \times 10^{-6} \moyr$~pc$^{-2}$) in the region between 200 and 300~pc. Between 300 and 500~pc, both $\Sigma_\mathrm{gas}$ and  $\Sigma_\mathrm{SFR}$ quickly decrease by more than one order of magnitude. The red bar highlights the region between $R=190$~pc and $R=310$~pc, corresponding to the possible range of values where the mass-weighted radius lies (we are excluding the central 5 pc for this calculation). This region fully overlaps the peak of the two distributions.
Therefore, disregarding the central few parsecs, most of the gas mass and star formation activity is located in a shell with thickness $\sim 100$~pc at $R\sim250$~pc, corresponding to the dense and star-forming CMZ ring/stream (see \autoref{CMZ}).

Finally, in the bottom panel of \autoref{Gas&SFRdistribution}, we plot the ratio between the time-averaged radial distributions of $\Sigma_\mathrm{gas}$ and $\Sigma_\mathrm{SFR}$, which roughly corresponds to the time-averaged radial distribution of depletion times.
The ratio $\langle\Sigma_\mathrm{gas}\rangle/\langle\Sigma_\mathrm{SFR}\rangle$ is low in the high density and star forming regions. It is $\sim10^8$~yr in the Galactic centre and in the shell between 200 and 300~pc, while it is almost $10^9$~yr elsewhere. Thus the ring from $\approx 200-300$~pc represents not only a local maximum of the gas surface density and the star formation rate, it represents a local minimum of the gas depletion time.

\subsection{Gas flows in and out of the CMZ}
\label{Flows}

For a better understanding of the gas cycle in the CMZ and its connection with the star formation cycle, we evaluate the mass flow rate across the cylindrical surface containing the CMZ. We separately calculate the mass flow rate across the lateral, $\dot{M}_\mathrm{R}$,  and top/bottom, $\dot{M}_\mathrm{z}$, surfaces of the cylinder ($R = 500$~pc, $h = 400$~pc).
The first one measures the net flow rate in the Galactic plane, the second one measures the flow rate orthogonal to the plane. We adopt a sign convention whereby positive values of $\dot{M}_\mathrm{R}$ and $\dot{M}_\mathrm{z}$ correspond to inflowing towards the CMZ.   

The left panel of \autoref{NetFlux} shows the time evolution of $\dot{M}_\mathrm{R}$ and $\dot{M}_\mathrm{z}$, color-coded in green and yellow, respectively. The instantaneous values of $\dot{M}_\mathrm{R}$ and $\dot{M}_\mathrm{z}$ rapidly vary, showing large fluctuations in a time window of a few Myr. In order to reduce this noise and to allow a clearer identification of an evolutionary trend, we smooth the instantaneous mass flow rates, convolving them with two 1D box kernels, one with width 10~Myr (dashed lines) and the other with width 100~Myr (solid lines).
$\dot{M}_\mathrm{R}$ displays large amplitude fluctuations when smoothed over a time period of 10~Myr, with values ranging from $-10\moyr$ to $10\moyr$. There are therefore short inflow/outflow cycles within the Galactic plane at the edge of the CMZ. Instead, when smoothed over a time period of 100~Myr, $\dot{M}_\mathrm{R}$ shows a roughly constant trend around $1-3 \moyr$, meaning that the gas inflow towards the CMZ is quasi-steady over longer time scales. While the short-period outflows/inflows are driven by local conditions of star formation and feedback, the continuous inflow of material depends on large-scale dynamics. It is interesting to note that the value of $\dot{M}_\mathrm{R}$ is consistent with the rate of mass inflow through the dust lanes ($\sim 2.7 \moyr$) estimated by \citet{Sormani+19}.

On the other side, $\dot{M}_\mathrm{z}$ shows weaker fluctuations regardless of the kernel width. Except for short time windows between 500 and 600~Myr,  $\dot{M}_\mathrm{z}$ is negative, meaning that the amount of gas escaping the Galactic disc is larger than the amount of gas falling back onto the disc. 
However, a trend as a function of time and SFR is clearly visible. When smoothed over a time period of $100$~Myr, $\dot{M}_\mathrm{z}$ increases from $-0.3\moyr$ at $t=200$~Myr, corresponding to a maximum in the star formation activity, to $-0.05 \moyr$ at $t=500$~Myr, corresponding to a minimum in the star formation activity (see top right panel of \autoref{SFR&GasMass}), and then decreases again. Therefore, higher star formation activity produces more powerful outflows.

\begin{figure*}
\includegraphics[width=\textwidth]{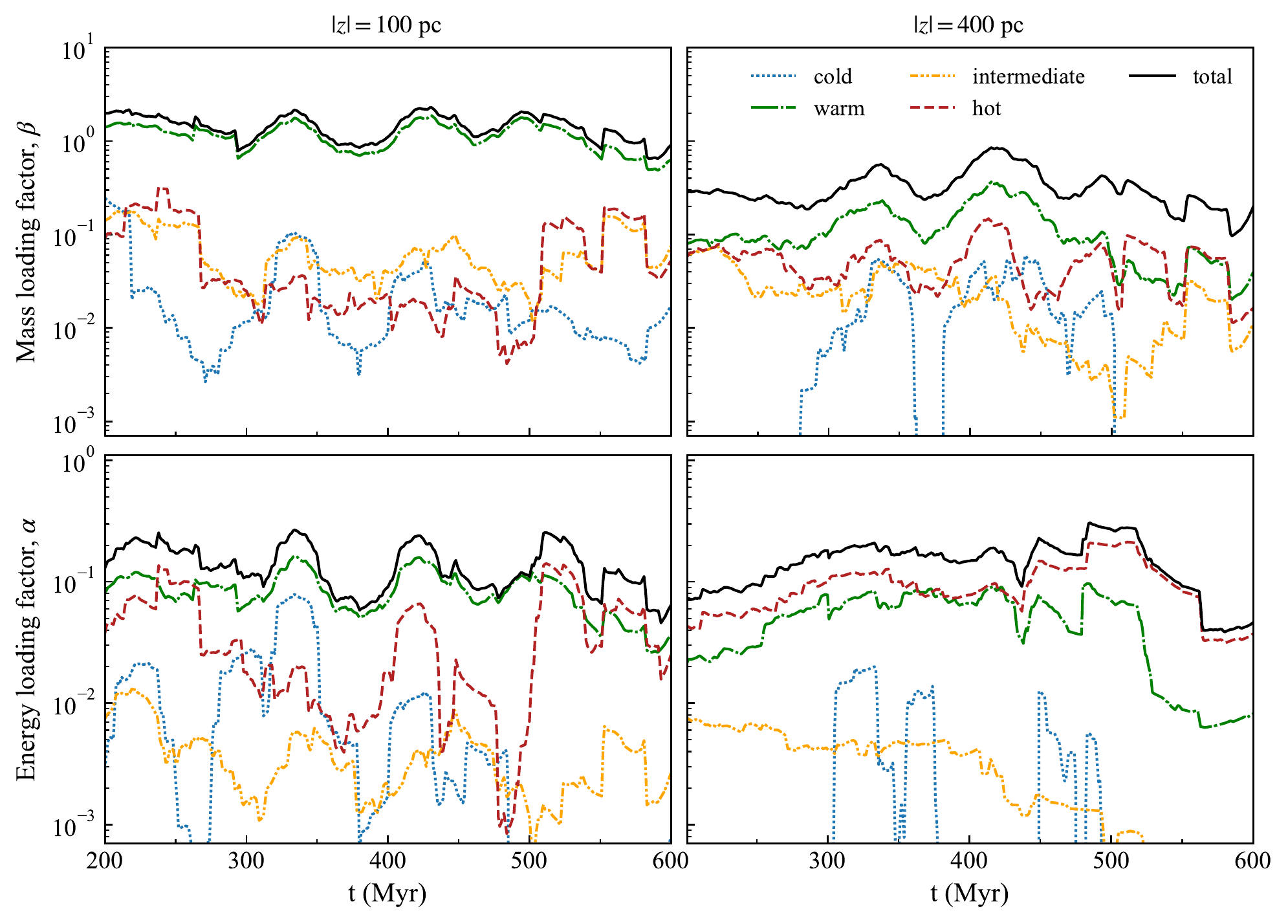}
\caption{Time evolution of mass (\textit{top panels}) and energy (\textit{bottom panels}) loading factors associated with outflowing gas at 100~pc (\textit{left panels}) and 400~pc (\textit{right panels}) above the Galactic plane in the CMZ region. The analysis has been done for thermally separated gas phases: cold ($T<5\times10^3$~K, green line), warm ($5\times10^3<T<2\times10^4$~K, green line), intermediate ($2\times10^4<T<5\times10^5$~K, yellow line) and hot ($T>5\times10^5$~K, brown line). The black lines indicate the total contribution from the three gas components.}
\label{LoadingFactors}
\end{figure*}

We note that $\dot{M}_\mathrm{R}$ is at most times larger than $\dot{M}_\mathrm{z}$, so there is a net gas mass inflow to the CMZ.
To evaluate whether the inflowing gas mass is converted into stars or accumulates within the CMZ, in the right panel of \autoref{NetFlux}, we show the total rate of variation in mass in the CMZ due to all sinks and sources, $\dot{M} = \mathrm{SFR} - \dot{M}_\mathrm{R} -\dot{M}_\mathrm{z}$, convolved with a 1D box kernel with width 100~Myr. The sign convection is that $\dot{M}$ is positive when the rate of star formation is larger than the net mass inflow rate, causing the total amount of gas mass to decrease. $\dot{M}$ exhibits a trend similar to the overall SFR (see top right panel of \autoref{SFR&GasMass}): in the time period between 200 and 300~Myr, when star formation activity is high ($\rm{SFR} \sim 2\moyr$), $\dot{M}$ is positive and gas is converted into stars before being replenished from the surroundings. In the time period between 300 and 500~Myr, when the star formation activity decreases, $\dot{M}$ is negative, meaning that mass is flowing into the CMZ faster than star formation can deplete it. In this time window, $\dot{M}$ exhibits fluctuations over the same time-scales of the SFR fluctuations, $~50-100$~Myr. 
After 500 Myr, as star formation activity increases, $\dot{M}$ increases until becoming positive again. Thus we find that the primary driver of variations in the SFR is not variations in the amount of mass in the CMZ, but instead variations in the efficiency with which this gas reservoir is converted to stars, which lead gas to either accumulate or deplete. However, all of these effects are relatively modest, so that the total mass in the CMZ only varies by tens of percent in time.

\subsection{Outflow loading factors}
\label{Outflow}

We now analyse the outflow activity driven by stellar feedback through the time evolution of two quantities, the mass and energy loading factors. The mass loading factor, $\beta$, is defined as
\begin{equation}
\beta = \dfrac{\dot{M}_\mathrm{z}^+}{\mathrm{SFR}}
\label{beta}
\end{equation}
where $\dot{M}_\mathrm{z}^+$ is the mass outflow rate along the vertical direction, $z$. The energy loading factor, $\alpha$, is defined as
\begin{equation}
\alpha = \dfrac{\dot{E}_\mathrm{Kin, z}^++\dot{E}_\mathrm{Th, z}^+}{E_\mathrm{SN} \,\mathrm{SFR}\,/ m_*}
\label{alpha}
\end{equation}
where $\dot{E}_\mathrm{Kin, z}^+$ and $\dot{E}_\mathrm{Th, z}^+$ are the kinetic and thermal energy outflow rates along $z$, respectively. Here $E_\mathrm{SN} \,\mathrm{SFR}\,/ m_*$ is the energy injection rate at the wind base, where $E_\mathrm{SN} =10^{51}$~erg is the total energy per supernova, and $m_* = 100\,\mo$ is the total mass of new stars per supernova.
The quantities $\dot{M}_\mathrm{z}^+$, $\dot{E}_\mathrm{Kin, z}^+$ and $\dot{E}_\mathrm{Th, z}^+$ are evaluated across circular areas with radius 500 pc located above and below the CMZ, at $|z|=100$~pc, corresponding to the CMZ scale height, and $|z|=400$~pc. We do not investigate regions at higher $|z|$ because in the halo region the gas density decreases and, given that our simulation has been run with a Meshless Finite Mass algorithm, the spatial resolution worsens. We include only outflowing gas in our calculations of $\dot{M}_\mathrm{z}^+$ and the analogous energy quantities, that is gas with $\mathrm{sgn}(v_\mathrm{z}) = \mathrm{sgn}(z)$. As in \autoref{SFR}, we convolve the mass and energy loading factors with a 1D box kernel with width 10~Myr, in order to reduce the high-frequency noise. 

\autoref{LoadingFactors} shows the evolution of mass (top panels) and energy (bottom panels) loading factors as a function of time at $|z|=100$~pc (left panels) and $|z|=400$~pc (right panels) for thermally separated gas phases. The cold phase (blue line) refers to gas at temperatures below $5\times10^3$~K. The warm phase (green line) refers to gas at temperatures between $5\times10^3$ and $2\times10^4$~K. The intermediate phase (yellow line) refers to gas at temperatures between $2\times10^4$ and $5\times10^5$~K. The hot phase (brown line) includes gas at temperatures larger than $5\times10^5$~K. In addition to the individual gas phases, we also calculate the contribution from the total gas (black solid line). 

The total mass loading factor substantially decreases with $|z|$. It is $\sim 1-2$ at $|z|=100$~pc, while it oscillates around 0.3 at $|z|=400$~pc, meaning that the outflowing mass flux carries off the CMZ $100-200\%$ of the gas mass converted into stars, but only $\sim 20 \%$ of it is able to travel larger distances and reach $|z|=400$~pc. Most of the outflowing gas falls back onto the disc acting as a fountain rather than as a wind. Focusing on the different thermal phases, we can note that the warm outflow largely dominates the mass flux and this causes the reduction of the total mass flux at higher $|z|$. Indeed, while the loading factor of the warm gas decreases by almost one order of magnitude from $|z|=100$~pc to $|z|=400$~pc, the hot- and intermediate-temperature gas exhibits roughly the same mass loading factor, between 0.01 and 0.1. The cold outflow weakly influences the final budget. It oscillates around 0.01 at $|z|=100$~pc, while it is completely absent in some time window at $|z|=400$~pc.

Unlike the mass loading factor, the total energy loading factor does not show significant variations with $|z|$. It presents larger fluctuations at $|z|=400$~pc, but the time-averaged value is similar at the two different heights. The average energy loading factor is $\sim 0.1-0.2$, meaning that the energy flux carries off  $10-20\%$ of the supernova energy budget. Most of energy has already been transferred to acceleration and heating of ambient gas or lost via radiative cooling. Different gas phases contribute in a different way at different $|z|$. At $z=100$~pc, in the region closer to the CMZ where the warm gas dominates in terms of mass, the warm outflow dominates the energy flux. The hot phase gives an important contribution to the total energy flux only in the time window of 200-300~Myr and 500-600~Myr, when the star formation activity is stronger.  The cold and intermediate gas phases weakly contribute to the total energy budget. At $z=400$~pc, the energy loading factor of the warm gas decreases, even though not to the same extent as the mass loading factor, while the hot outflow dominates the energy flux. The energy loading factor of the hot gas increases with $|z|$, going from a few~$\times\;0.01$ at $|z|=100$~pc to $\sim 0.1$ at $|z|=400$~pc. This increase of energy might be due either to accretion of hot halo gas accelerated by the outflow or to intermediate-temperature gas that evaporates in the hotter outflow, thus increasing its mass and energy. The same trend is mildly visible in the mass loading factor plots, where the average mass loading factor of the hot gas slightly increases at higher $|z|$, while the average mass loading factor of the intermediate-temperature gas slightly decreases.

\section{Discussion}
\label{Discussion}

\begin{figure}
\includegraphics[width=0.49\textwidth]{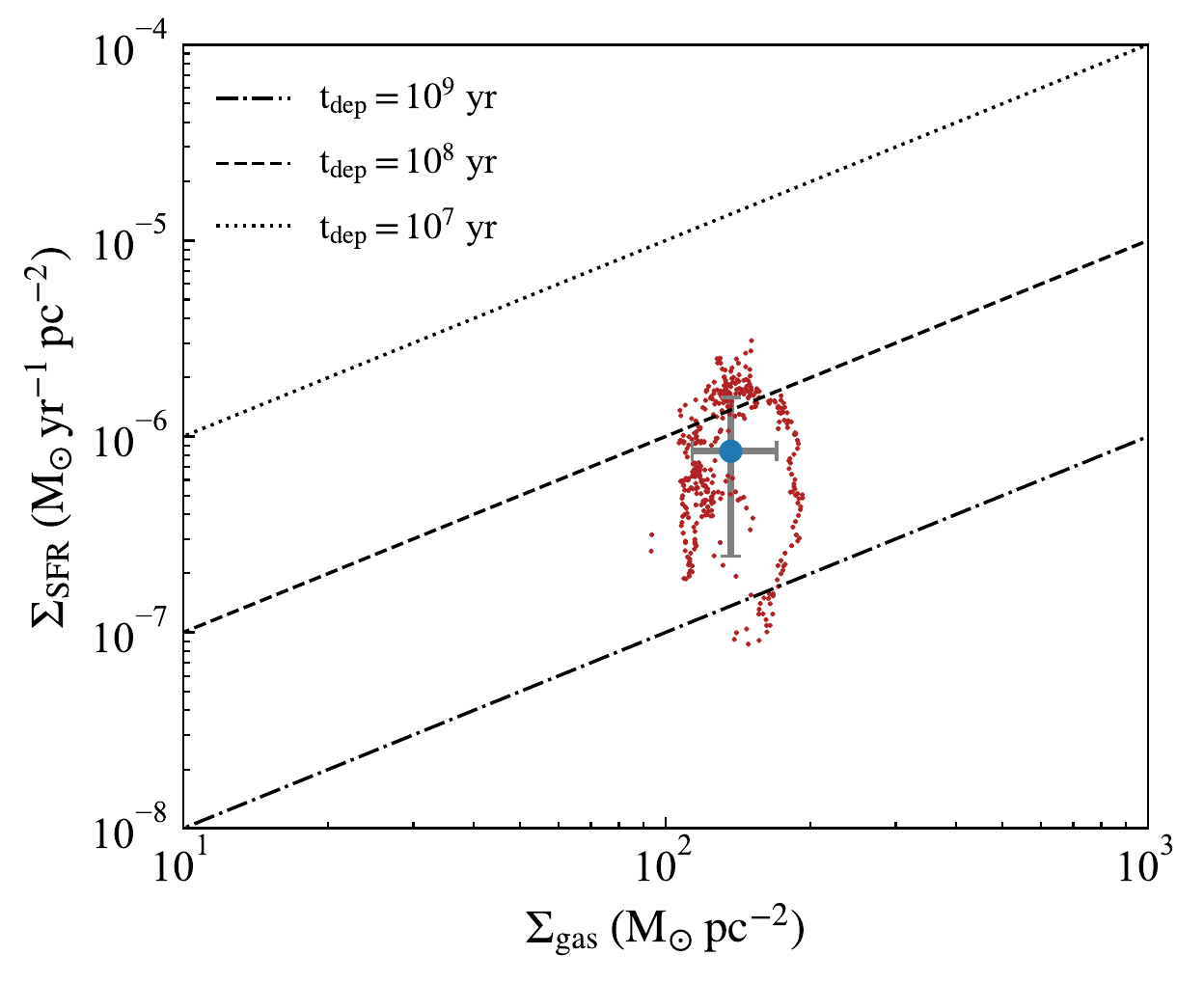}
\caption{Kennicutt-Schmidt plot showing SFR surface density, $\Sigma_\mathrm{SFR}$, versus gas surface density, $\Sigma_\mathrm{gas}$. The red dots represent the distribution of the simulation outcomes in the CMZ, sampled every 1~Myr. The blue dot indicates the average of the distribution, while the grey bands are the 16 and 84 percentiles along each direction. The black lines show the $\Sigma_\mathrm{gas}-\Sigma_\mathrm{SFR}$ relation for constant depletion times of $10^7$~yr (dotted line), $10^8$~yr (dashed line) and $10^9$~yr (dash-dotted line).}
\label{KSPlot}
\end{figure}

\subsection{Star formation in nuclear regions}

In \autoref{KSPlot}, we show the results of our simulation (red dots) in the $\Sigma_\mathrm{SFR}-\Sigma_\mathrm{gas}$ Kennicutt-Schmidt \citep{Schmidt59,Kennicutt98} plot. SFR surface densities are calculated dividing the SFR averaged over a 10~Myr time window (top right panel of \autoref{SFR&GasMass}) by the area of the CMZ, $\pi R^2$, where $R=500$~pc. Gas surface densities are calculated using the instantaneous mass within the CMZ (top left panel of \autoref{SFR&GasMass}) divided by $\pi R^2$. The black lines show the $\Sigma_\mathrm{gas}-\Sigma_\mathrm{SFR}$ relation for constant values of $t_\mathrm{dep}$. For reference, the depletion time that characterizes star formation in the outer regions of spiral galaxies is $\sim 2\times10^9$~yr \citep[e.g.][]{Bigiel+08,Leroy+13,Utomo+17}. The blue point indicates the average of the $\Sigma_\mathrm{SFR}-\Sigma_\mathrm{gas}$ distribution in the simulation, while the grey bands indicate the 16 and 84 percentiles. The simulation data points present a large scatter along the vertical direction, and are clearly far away from tracing a relation based on a constant depletion time (see also \autoref{SFR}), which instead holds for `normal' galaxy discs. It is interesting to note that the scatter in $\Sigma_\mathrm{gas}$ is lower than a factor 2, while the scatter in $\Sigma_\mathrm{SFR}$ spans over almost one order of magnitude. This result leads to the same conclusion we inferred in \autoref{Flows}. Variations in the SFR are caused by variations in the star formation efficiency, rather that changes in the amount of gas mass within the CMZ. Moreover, most of the dataset is below the dash-dotted line, corresponding to $t_\mathrm{dep} = 10^9$~yr, suggesting that the CMZ might be more efficient at forming stars per unit gas mass than normal galaxy discs.

The large scatter in $\Sigma_\mathrm{SFR}$ is a consequence of the cyclic nature of the SFR. Since the characteristic time-scale for star-formation variability, $\sim50$~Myr, is comparable to the time-scale for a population of young massive stars to blow out, we identify the following mechanism to explain such cyclic state.
In the highly-dense CMZ, the dynamical times of star formation ($1-2$~Myr) - both the free-fall time of individual molecular clouds and the orbital time of the entire CMZ - are about one order of magnitude smaller than the time required for stellar feedback to be effective (tens of Myr). Thus, gas keeps forming stars at a high rate until feedback from young stars raises the level of turbulence and significantly suppresses star formation. In turn, the low level of star formation entails a low level of feedback, which prevents the turbulence dissipation, causing the gas to collapse and restarting the cycle.  

Indications of episodic star formation in galactic nuclei have been found in other simulations of Milky Way-like galaxies \citep[e.g.][]{Emsellem+15, Torrey+17, Seo+19}. Among them, \citet{Torrey+17} reached a conclusion similar to ours, i.e. stellar feedback cannot immediately balance the rapid star formation, thus leading to burst/quench cycle of star formation. However, in their simulations, the main cause for the variations in the SFR is the expulsion/accumulation of gas in the nuclear region, rather than the variations in the star formation efficiency. We find the opposite: the gas mass in the CMZ stays nearly constant, and variations are driven by changes in the level of turbulence, as proposed by \citet{Krumholz+17}. Most likely, this disagreement is due to the fact that \citet{Torrey+17} did not account for the effect of a galactic bar, which causes gas from the outer parts of the disc to continuously flow towards the CMZ with a rate comparable to or larger than the instantaneous SFR. In the absence of such a continuous mass supply, it should be much easier to evacuate the CMZ. 

Unlike \citet{Torrey+17}, \citet{Seo+19} performed simulations in the presence of a galactic bar. They concluded that the SFR well correlates with the mass inflow rate into the CMZ. We point out that their conclusion might seem in contrast with our previous statement, i.e. SFR variations are not primarily driven by variations in the CMZ mass. However, the correlation found by \citet{Seo+19} holds on timescales of $\sim 1$~Gyr, while it ceases on timescales of tens/hundreds of Myr \citep[see Figure 16a in][]{Seo+19}, similar to what we find in our work. At this stage, we cannot speculate whether we should see a similar correlation if we ran our simulation for a few Gyr. An important aspect that might make the difference in the long-term evolution of the Galaxy is that in \citet{Seo+19} bars are formed via stellar dynamical effects, rather than being tuned to match the observed Milky Way potential as in our work.

Beyond the Milky-Way's CMZ, our results on the cyclic nature of star formation might be extended to similar nuclear regions in barred spiral galaxies. Indeed, modellings of the optical spectra of some nearby galaxies have suggested that their nuclear regions might be characterized by episodic bursts of star formation activity \citep{Sarzi+07}. Moreover, observational findings have shown that, on average, the nuclear regions of star-forming galaxies exhibit a much wider range of depletion times than the outer discs \citep{Leroy+13}.
The physics underneath should be the same. Due to the presence of a rotating bar, gas flows towards the galactic centre forming a highly dense and star-forming ring where the extreme environmental conditions do not allow star formation to reach a steady-state feedback-regulated equilibrium. Thus, single data points in \autoref{KSPlot} might represent different evolutionary snapshots of the Milky-Way's CMZ, as well as a random sample of nuclear regions in different barred spiral galaxies, each of them in a different stage of the star formation cycle. Indeed, the distribution of depletion times shown in \autoref{KSPlot} is in excellent agreement with the distribution observed in the centres of nearby spiral galaxies by \citet[their figure 13]{Leroy+13}, who find a range from $t_{\rm dep} \approx 2\times 10^9$ yr for the most inefficient CMZs to $t_{\rm dep} \approx 10^8$ yr for the most active ones.

\subsection{Supernova-driven outflow in the CMZ}

The outcomes of our simulation have shown that stellar feedback drives a mainly two-phase wind off the CMZ. The warm phase largely dominates the mass flux in the region immediately outside the Galactic disc, carrying off $100-200\%$ of the gas mass converted into stars. At larger heights above the plane the contribution of the warm gas decreases because part of this gas falls back onto the disc. At $|z|=400$~pc, the mass loading factor of the warm gas is reduced by almost one order of magnitude. The hot phase dominates the energy flux in the Galactic halo, where it carries off $\sim 10\%$ of the supernova energy budget. Unlike the warm phase, the hot mass flux does not exhibit significant variations with $|z|$, suggesting that the hot component of the Galactic outflow might travel large distances in the Galactic halo. However, we can not make accurate predictions since we limit our analysis at $|z|\leq400$~pc. 

These results are overall consistent with the outcomes of recent kpc-scale simulations exploring the properties of supernova-driven outflows in different galactic environments. \citet{Kim&Ostriker18} investigated the outflow properties in the solar neighbourhood from $|z|=0$ to $|z|=4$~kpc. They also found that the outflowing gas mainly consists of warm fountain flows, which dominate the mass flux close to the midplane, and hot gas, potentially able to escape the Galaxy potential. However, it is difficult to make a direct comparison with our work since \citet{Kim&Ostriker18} explored a different Milky-Way environment, less dense ($\Sigma_\mathrm{gas} \sim 10\,\mo\mathrm{pc}^{-2}$) and with a weaker gravitational potential ($h=400$~pc). If we compare our results at $|z|=400$~pc with theirs at $|z|=1$~kpc, we note two main differences. First, the mass loading factor of the warm gas is almost one order of magnitude higher in \citeauthor{Kim&Ostriker18}'s simulation. Second, they found that hot gas largely dominates the outflow energetics. A decrease of mass loading factor in extremely dense environments ($\Sigma_\mathrm{gas} \gtrsim 10^2\,\mo\mathrm{pc}^{-2}$) has been confirmed in other works \citep[e.g.][]{Martizzi+16, Li+17}. However, in agreement with \citet{Kim&Ostriker18}, \citet{Li+17} found that at $|z|\gtrsim1$~kpc the hot phase completely dominates the energy flux. At this stage, we cannot determine whether the higher proportion of energy carried by the warm phase in our simulations is due to the fact that we are simulating an environment where there is a far larger amount of warm and cold gas overall, or if we underestimate the real amount of hot gas because we cannot resolve the cooling radius in high-density regions ($n>10-100$~cm$^{-3}$). 
If we focus on the loading factor of the hot gas only, $\simeq 0.1$, we can anyway conclude that such value is in agreement with the one inferred by \citet{Li+17} in highly dense environments.

\subsubsection{Comparison with observations}

We now derive some quantitative properties of the present-day supernova-driven nuclear outflow of the Milky Way, given the observed SFR of the CMZ. Since the evolution of mass and energy loading factors does not present significant variations as a function of time/SFR at $|z|=400$~pc, in the following we use time-averaged quantities.
Our simulation shows that at $|z|=400$~pc the average mass and energy loading factors are $\langle\beta\rangle \simeq 0.3$ and $\langle\alpha\rangle \simeq 0.15$, respectively. Given a SFR of $0.04-0.1\,\mo$, the mass outflow rate is $\simeq 1-3\times10^{-2}\,\moyr$, while the energy injection rate is $\simeq 1-4\times10^{39}$~erg~s$^{-1}$ (see \autoref{beta} and \autoref{alpha}). Focusing on the two main thermal phases, the mass and energy outflow rates of the warm component are $\simeq 0.5-1.5\times 10^{-2}\,\moyr$ and $\simeq 0.3-1\times10^{39}$~erg~s$^{-1}$, respectively ($\langle\beta\rangle\simeq 0.15$ and $\langle\alpha\rangle\simeq 0.04$), while the mass and energy outflow rates of the hot component are $\simeq 0.2-0.6\times10^{-2} \,\moyr $ and $\simeq 0.6-3\times10^{39}$~erg~s$^{-1}$, respectively ($\langle\beta\rangle\simeq 0.06$ and $\langle\alpha\rangle\simeq 0.1$).

By modelling the \hi\ Galactic nuclear outflow within 1.5~kpc above the disc, \citet{DiTeodoro+18} estimated that the \hi\ mass outflow rate is $\sim 0.1\moyr$, while the kinetic energy injection rate is $\sim 5 \times10^{39}$~erg~s$^{-1}$. These values are significantly higher that the ones inferred by the simulation for the warm component of the outflow. The disagreement may be due either to uncertainties in the models used to obtain the observational estimates, or to some physical process that we do not account for in our simulation. Indeed, star formation activity in the CMZ might not be the sole source of the Galactic nuclear outflow. For example, using \ovii\ an \oviii\ line detections, \citet{Miller&Bregman16} modelled the Fermi Bubbles as a galactic outflow going through Sedov-Taylor expansion and inferred an energy injection rate of $\sim 2.3\times10^{42}$~erg~s$^{-1}$. This value is clearly inconsistent with the energetics of an outflow powered by supernova feedback, while it is consistent with the energetics of a bubble generated by an AGN wind. 

\section{Conclusions}
\label{Conclusions}

The CMZ represents an extreme environment in our Galaxy, characterized by densities orders of magnitude larger than those measured in the outer disc, but a level of star formation significantly below the one predicted by currently accepted SFR prescriptions. In this paper, we present a detailed three-dimensional hydrodynamical simulation aiming at unravelling the gas cycle and star formation history in the innermost region of a Milky Way-like barred spiral galaxy. The major findings of our work are as follows:
\begin{itemize}
\item Due to the presence of the non-axisymmetric bar gravitational potential, gas in the outer part of the disc slowly drifts towards the CMZ forming a highly-dense ($\Sigma_\mathrm{gas} \gtrsim10^2 \mo$) stream/ring at $\approx 200-300$~pc from the Galactic centre. Within the inner 500~pc region of the Galaxy, this ring represents not only a maximum of the star formation rate, but also a minimum of the gas depletion time - maximum of the star formation efficiency (see \autoref{Gas&SFRdistribution}).
\item Star formation activity in the CMZ goes through oscillatory burst-quench cycles (see \autoref{SFR&GasMass}), with characteristic variability-time of $50$~Myr mainly driven by feedback instabilities. The dense CMZ ring undergoes bursts of star formation over dynamical times (a few Myr) shorter than the lifetimes of massive stars (tens of Myr). This leads to an unstable feedback-regulated system with alternating cycles of bursts and suppressions in star formation activity. A comparison with the present-day SFR of CMZ suggests that it might lie at a minimum of a longer star formation period.
\item Throughout the simulation time, there is a quasi-steady net inflow of gas towards the CMZ ($1-3\moyr$, see \autoref{NetFlux}), and the total gas mass in the CMZ remains constant to within a few tens of percent. Thus, variations in the SFR cannot be primarily due to accretion/expulsion of gas  in the CMZ region, but instead to large oscillations in the instantaneous gas depletion time/star formation efficiency (see \autoref{SFR&GasMass}) due to periodic variations in the level of supernova-driven turbulence. The range of depletion times produced in our simulation agrees well with the distribution observed in the central regions of nearby spiral galaxies.
\item Supernova feedback drives a mainly two-phase wind off the CMZ (see \autoref{LoadingFactors}). The warm phase largely dominates the mass flux near the Galactic plane, carrying $100-200\%$ of the gas mass converted into stars. However, most of this gas behaves like a fountain flow, falling back onto the disc rather than escaping the Galaxy potential. The hot phase carries most of the energy, with an energy injection rate equal to $10-20\%$ of the supernova energy budget.
\end{itemize}
These results might be relevant to explain the observational properties of the CMZ and, more in general, of nuclear regions of Milky Way-like galaxies. Combining these theoretical predictions with observational findings might have important implications in characterizing the evolutionary path of these extreme galactic environments.

\section*{Acknowledgements}
The authors thank M\`{e}lanie Chevance, Roland Crocker, Chang-Goo Kim, Diederik Kruijssen, Mattia Sormani and Robin Tre{\ss} for useful discussions and suggestions. LA thanks Eric Gentry for his initial guidance in using \textsc{GIZMO}. The simulation was performed on the Raijin supercomputer at the National Computational Infrastructure (NCI), which is supported by the Australian Government, through grant jh2. LA and MRK acknowledge support from the Australian Research Council's \textit{Discovery Projects} and \textit{Future Fellowships} funding schemes, awards DP190101258 and FT180100375. EDT and NMG acknowledge the support of the Australian Research Council (ARC) through grant DP160100723. NMG acknowledges the support of the ARC through \textit{Future Fellowships} FT150100024.




\bibliographystyle{mnras}
\bibliography{biblio}

\begin{thebibliography}{}
\makeatletter
\relax
\def\mn@urlcharsother{\let\do\@makeother \do\$\do\&\do\#\do\^\do\_\do\%\do\~}
\def\mn@doi{\begingroup\mn@urlcharsother \@ifnextchar [ {\mn@doi@}
  {\mn@doi@[]}}
\def\mn@doi@[#1]#2{\def\@tempa{#1}\ifx\@tempa\@empty \href
  {http://dx.doi.org/#2} {doi:#2}\else \href {http://dx.doi.org/#2} {#1}\fi
  \endgroup}
\def\mn@eprint#1#2{\mn@eprint@#1:#2::\@nil}
\def\mn@eprint@arXiv#1{\href {http://arxiv.org/abs/#1} {{\tt arXiv:#1}}}
\def\mn@eprint@dblp#1{\href {http://dblp.uni-trier.de/rec/bibtex/#1.xml}
  {dblp:#1}}
\def\mn@eprint@#1:#2:#3:#4\@nil{\def\@tempa {#1}\def\@tempb {#2}\def\@tempc
  {#3}\ifx \@tempc \@empty \let \@tempc \@tempb \let \@tempb \@tempa \fi \ifx
  \@tempb \@empty \def\@tempb {arXiv}\fi \@ifundefined
  {mn@eprint@\@tempb}{\@tempb:\@tempc}{\expandafter \expandafter \csname
  mn@eprint@\@tempb\endcsname \expandafter{\@tempc}}}

\bibitem[\protect\citeauthoryear{{Barnes}, {Longmore}, {Battersby}, {Bally},
  {Kruijssen}, {Henshaw}  \& {Walker}}{{Barnes} et~al.}{2017}]{Barnes+17}
{Barnes} A.~T.,  {Longmore} S.~N.,  {Battersby} C.,  {Bally} J.,  {Kruijssen}
  J.~M.~D.,  {Henshaw} J.~D.,   {Walker} D.~L.,  2017, \mn@doi [\mnras]
  {10.1093/mnras/stx941}, \href
  {http://cdsads.u-strasbg.fr/abs/2017MNRAS.469.2263B} {469, 2263}

\bibitem[\protect\citeauthoryear{{Battersby}, {Bally}  \&
  {Svoboda}}{{Battersby} et~al.}{2017}]{Battersby+17}
{Battersby} C.,  {Bally} J.,   {Svoboda} B.,  2017, \mn@doi [\apj]
  {10.3847/1538-4357/835/2/263}, \href
  {http://cdsads.u-strasbg.fr/abs/2017ApJ...835..263B} {835, 263}

\bibitem[\protect\citeauthoryear{{Bigiel}, {Leroy}, {Walter}, {Brinks}, {de
  Blok}, {Madore}  \& {Thornley}}{{Bigiel} et~al.}{2008}]{Bigiel+08}
{Bigiel} F.,  {Leroy} A.,  {Walter} F.,  {Brinks} E.,  {de Blok} W.~J.~G.,
  {Madore} B.,   {Thornley} M.~D.,  2008, \mn@doi [\aj]
  {10.1088/0004-6256/136/6/2846}, \href
  {http://cdsads.u-strasbg.fr/abs/2008AJ....136.2846B} {136, 2846}

\bibitem[\protect\citeauthoryear{{Binney} \& {Tremaine}}{{Binney} \&
  {Tremaine}}{2008}]{Binney&Tremaine08}
{Binney} J.,  {Tremaine} S.,  2008, {Galactic Dynamics: Second Edition}.
Princeton University Press

\bibitem[\protect\citeauthoryear{{Binney}, {Gerhard}, {Stark}, {Bally}  \&
  {Uchida}}{{Binney} et~al.}{1991}]{Binney+91}
{Binney} J.,  {Gerhard} O.~E.,  {Stark} A.~A.,  {Bally} J.,   {Uchida} K.~I.,
  1991, \mn@doi [\mnras] {10.1093/mnras/252.2.210}, \href
  {http://cdsads.u-strasbg.fr/abs/1991MNRAS.252..210B} {252, 210}

\bibitem[\protect\citeauthoryear{{Bissantz} \& {Gerhard}}{{Bissantz} \&
  {Gerhard}}{2002}]{Bissantz&Gerhard02}
{Bissantz} N.,  {Gerhard} O.,  2002, \mn@doi [\mnras]
  {10.1046/j.1365-8711.2002.05116.x}, \href
  {http://cdsads.u-strasbg.fr/abs/2002MNRAS.330..591B} {330, 591}

\bibitem[\protect\citeauthoryear{{Bland-Hawthorn} \& {Cohen}}{{Bland-Hawthorn}
  \& {Cohen}}{2003}]{Bland-Hawthorn&Cohen03}
{Bland-Hawthorn} J.,  {Cohen} M.,  2003, \mn@doi [\apj] {10.1086/344573}, \href
  {http://cdsads.u-strasbg.fr/abs/2003ApJ...582..246B} {582, 246}

\bibitem[\protect\citeauthoryear{{Bressan}, {Marigo}, {Girardi}, {Salasnich},
  {Dal Cero}, {Rubele}  \& {Nanni}}{{Bressan} et~al.}{2012}]{Bressan+12}
{Bressan} A.,  {Marigo} P.,  {Girardi} L.,  {Salasnich} B.,  {Dal Cero} C.,
  {Rubele} S.,   {Nanni} A.,  2012, \mn@doi [\mnras]
  {10.1111/j.1365-2966.2012.21948.x}, \href
  {http://adsabs.harvard.edu/abs/2012MNRAS.427..127B} {427, 127}

\bibitem[\protect\citeauthoryear{{Chabrier}}{{Chabrier}}{2005}]{Chabrier05}
{Chabrier} G.,  2005, in {Corbelli} E.,  {Palla} F.,   {Zinnecker} H.,  eds,
  Astrophysics and Space Science Library Vol. 327, The Initial Mass Function 50
  Years Later. p.~41 (\mn@eprint {} {astro-ph/0409465}),
  \mn@doi{10.1007/978-1-4020-3407-7_5}

\bibitem[\protect\citeauthoryear{{Dekel} \& {Krumholz}}{{Dekel} \&
  {Krumholz}}{2013}]{Dekel+13}
{Dekel} A.,  {Krumholz} M.~R.,  2013, \mn@doi [\mnras] {10.1093/mnras/stt480},
  \href {http://adsabs.harvard.edu/abs/2013MNRAS.432..455D} {432, 455}

\bibitem[\protect\citeauthoryear{{Dekel}, {Sari}  \& {Ceverino}}{{Dekel}
  et~al.}{2009}]{Dekel+09}
{Dekel} A.,  {Sari} R.,   {Ceverino} D.,  2009, \mn@doi [\apj]
  {10.1088/0004-637X/703/1/785}, \href
  {http://adsabs.harvard.edu/abs/2009ApJ...703..785D} {703, 785}

\bibitem[\protect\citeauthoryear{{Di Teodoro}, {McClure-Griffiths}, {Lockman},
  {Denbo}, {Endsley}, {Ford}  \& {Harrington}}{{Di Teodoro}
  et~al.}{2018}]{DiTeodoro+18}
{Di Teodoro} E.~M.,  {McClure-Griffiths} N.~M.,  {Lockman} F.~J.,  {Denbo}
  S.~R.,  {Endsley} R.,  {Ford} H.~A.,   {Harrington} K.,  2018, \mn@doi [\apj]
  {10.3847/1538-4357/aaad6a}, \href
  {http://cdsads.u-strasbg.fr/abs/2018ApJ...855...33D} {855, 33}

\bibitem[\protect\citeauthoryear{{Emsellem}, {Renaud}, {Bournaud}, {Elmegreen},
  {Combes}  \& {Gabor}}{{Emsellem} et~al.}{2015}]{Emsellem+15}
{Emsellem} E.,  {Renaud} F.,  {Bournaud} F.,  {Elmegreen} B.,  {Combes} F.,
  {Gabor} J.~M.,  2015, \mn@doi [\mnras] {10.1093/mnras/stu2209}, \href
  {http://adsabs.harvard.edu/abs/2015MNRAS.446.2468E} {446, 2468}

\bibitem[\protect\citeauthoryear{{Ferland} et~al.,}{{Ferland}
  et~al.}{2013}]{Ferland+13}
{Ferland} G.~J.,  et~al., 2013, \rmxaa, \href
  {http://cdsads.u-strasbg.fr/abs/2013RMxAA..49..137F} {49, 137}

\bibitem[\protect\citeauthoryear{{Ferri{\`e}re}, {Gillard}  \&
  {Jean}}{{Ferri{\`e}re} et~al.}{2007}]{Ferriere+07}
{Ferri{\`e}re} K.,  {Gillard} W.,   {Jean} P.,  2007, \mn@doi [\aap]
  {10.1051/0004-6361:20066992}, \href
  {http://cdsads.u-strasbg.fr/abs/2007A%26A...467..611F} {467, 611}

\bibitem[\protect\citeauthoryear{{Fox} et~al.,}{{Fox} et~al.}{2015}]{Fox+15}
{Fox} A.~J.,  et~al., 2015, \mn@doi [\apjl] {10.1088/2041-8205/799/1/L7}, \href
  {http://cdsads.u-strasbg.fr/abs/2015ApJ...799L...7F} {799, L7}

\bibitem[\protect\citeauthoryear{{Gentry}, {Krumholz}, {Madau}  \&
  {Lupi}}{{Gentry} et~al.}{2019}]{Gentry+19}
{Gentry} E.~S.,  {Krumholz} M.~R.,  {Madau} P.,   {Lupi} A.,  2019, \mn@doi
  [\mnras] {10.1093/mnras/sty3319}, \href
  {http://adsabs.harvard.edu/abs/2019MNRAS.483.3647G} {483, 3647}

\bibitem[\protect\citeauthoryear{{Gilmore} \& {Reid}}{{Gilmore} \&
  {Reid}}{1983}]{Gilmore&Reid83}
{Gilmore} G.,  {Reid} N.,  1983, \mn@doi [\mnras] {10.1093/mnras/202.4.1025},
  \href {http://cdsads.u-strasbg.fr/abs/1983MNRAS.202.1025G} {202, 1025}

\bibitem[\protect\citeauthoryear{{Ginsburg} et~al.,}{{Ginsburg}
  et~al.}{2016}]{Ginsburg+16}
{Ginsburg} A.,  et~al., 2016, \mn@doi [\aap] {10.1051/0004-6361/201526100},
  \href {http://cdsads.u-strasbg.fr/abs/2016A%26A...586A..50G} {586, A50}

\bibitem[\protect\citeauthoryear{{Glassgold}, {Galli}  \&
  {Padovani}}{{Glassgold} et~al.}{2012}]{Glassgold12a}
{Glassgold} A.~E.,  {Galli} D.,   {Padovani} M.,  2012, \mn@doi [\apj]
  {10.1088/0004-637X/756/2/157}, \href
  {http://adsabs.harvard.edu/abs/2012ApJ...756..157G} {756, 157}

\bibitem[\protect\citeauthoryear{{Henshaw} et~al.,}{{Henshaw}
  et~al.}{2016}]{Henshaw+16}
{Henshaw} J.~D.,  et~al., 2016, \mn@doi [\mnras] {10.1093/mnras/stw121}, \href
  {http://cdsads.u-strasbg.fr/abs/2016MNRAS.457.2675H} {457, 2675}

\bibitem[\protect\citeauthoryear{{Heyer}, {Gutermuth}, {Urquhart}, {Csengeri},
  {Wienen}, {Leurini}, {Menten}  \& {Wyrowski}}{{Heyer}
  et~al.}{2016}]{Heyer+16}
{Heyer} M.,  {Gutermuth} R.,  {Urquhart} J.~S.,  {Csengeri} T.,  {Wienen} M.,
  {Leurini} S.,  {Menten} K.,   {Wyrowski} F.,  2016, \mn@doi [\aap]
  {10.1051/0004-6361/201527681}, \href
  {http://cdsads.u-strasbg.fr/abs/2016A%26A...588A..29H} {588, A29}

\bibitem[\protect\citeauthoryear{{Hopkins}}{{Hopkins}}{2015}]{Hopkins15}
{Hopkins} P.~F.,  2015, \mn@doi [\mnras] {10.1093/mnras/stv195}, \href
  {http://cdsads.u-strasbg.fr/abs/2015MNRAS.450...53H} {450, 53}

\bibitem[\protect\citeauthoryear{{Hopkins} et~al.,}{{Hopkins}
  et~al.}{2018a}]{Hopkins+18b}
{Hopkins} P.~F.,  et~al., 2018a, \mn@doi [\mnras] {10.1093/mnras/sty674}, \href
  {http://cdsads.u-strasbg.fr/abs/2018MNRAS.477.1578H} {477, 1578}

\bibitem[\protect\citeauthoryear{{Hopkins} et~al.,}{{Hopkins}
  et~al.}{2018b}]{Hopkins+18a}
{Hopkins} P.~F.,  et~al., 2018b, \mn@doi [\mnras] {10.1093/mnras/sty1690},
  \href {http://cdsads.u-strasbg.fr/abs/2018MNRAS.480..800H} {480, 800}

\bibitem[\protect\citeauthoryear{{Immer}, {Schuller}, {Omont}  \&
  {Menten}}{{Immer} et~al.}{2012}]{Immer+12}
{Immer} K.,  {Schuller} F.,  {Omont} A.,   {Menten} K.~M.,  2012, \mn@doi
  [\aap] {10.1051/0004-6361/201117857}, \href
  {http://cdsads.u-strasbg.fr/abs/2012A%26A...537A.121I} {537, A121}

\bibitem[\protect\citeauthoryear{{Kennicutt}}{{Kennicutt}}{1998}]{Kennicutt98}
{Kennicutt} Jr. R.~C.,  1998, \mn@doi [\apj] {10.1086/305588}, \href
  {http://cdsads.u-strasbg.fr/abs/1998ApJ...498..541K} {498, 541}

\bibitem[\protect\citeauthoryear{{Kim} \& {Ostriker}}{{Kim} \&
  {Ostriker}}{2015}]{Kim&Ostriker15}
{Kim} C.-G.,  {Ostriker} E.~C.,  2015, \mn@doi [\apj]
  {10.1088/0004-637X/802/2/99}, \href
  {http://adsabs.harvard.edu/abs/2015ApJ...802...99K} {802, 99}

\bibitem[\protect\citeauthoryear{{Kim} \& {Ostriker}}{{Kim} \&
  {Ostriker}}{2018}]{Kim&Ostriker18}
{Kim} C.-G.,  {Ostriker} E.~C.,  2018, \mn@doi [\apj]
  {10.3847/1538-4357/aaa5ff}, \href
  {http://cdsads.u-strasbg.fr/abs/2018ApJ...853..173K} {853, 173}

\bibitem[\protect\citeauthoryear{{Kim} \& {Stone}}{{Kim} \&
  {Stone}}{2012}]{Kim&Stone12}
{Kim} W.-T.,  {Stone} J.~M.,  2012, \mn@doi [\apj]
  {10.1088/0004-637X/751/2/124}, \href
  {http://cdsads.u-strasbg.fr/abs/2012ApJ...751..124K} {751, 124}

\bibitem[\protect\citeauthoryear{{Kim}, {Seo}, {Stone}, {Yoon}  \&
  {Teuben}}{{Kim} et~al.}{2012}]{Kim+12}
{Kim} W.-T.,  {Seo} W.-Y.,  {Stone} J.~M.,  {Yoon} D.,   {Teuben} P.~J.,  2012,
  \mn@doi [\apj] {10.1088/0004-637X/747/1/60}, \href
  {http://cdsads.u-strasbg.fr/abs/2012ApJ...747...60K} {747, 60}

\bibitem[\protect\citeauthoryear{{Kruijssen} \& {Longmore}}{{Kruijssen} \&
  {Longmore}}{2013}]{Kruijssen&Longmore13}
{Kruijssen} J.~M.~D.,  {Longmore} S.~N.,  2013, \mn@doi [\mnras]
  {10.1093/mnras/stt1634}, \href
  {http://cdsads.u-strasbg.fr/abs/2013MNRAS.435.2598K} {435, 2598}

\bibitem[\protect\citeauthoryear{{Kruijssen}, {Longmore}, {Elmegreen},
  {Murray}, {Bally}, {Testi}  \& {Kennicutt}}{{Kruijssen}
  et~al.}{2014}]{Kruijssen+14}
{Kruijssen} J.~M.~D.,  {Longmore} S.~N.,  {Elmegreen} B.~G.,  {Murray} N.,
  {Bally} J.,  {Testi} L.,   {Kennicutt} R.~C.,  2014, \mn@doi [\mnras]
  {10.1093/mnras/stu494}, \href
  {http://cdsads.u-strasbg.fr/abs/2014MNRAS.440.3370K} {440, 3370}

\bibitem[\protect\citeauthoryear{{Kruijssen}, {Dale}  \&
  {Longmore}}{{Kruijssen} et~al.}{2015}]{Kruijssen+15}
{Kruijssen} J.~M.~D.,  {Dale} J.~E.,   {Longmore} S.~N.,  2015, \mn@doi
  [\mnras] {10.1093/mnras/stu2526}, \href
  {http://cdsads.u-strasbg.fr/abs/2015MNRAS.447.1059K} {447, 1059}

\bibitem[\protect\citeauthoryear{{Krumholz} \& {Gnedin}}{{Krumholz} \&
  {Gnedin}}{2011}]{Krumholz&Gnedin11}
{Krumholz} M.~R.,  {Gnedin} N.~Y.,  2011, \mn@doi [\apj]
  {10.1088/0004-637X/729/1/36}, \href
  {http://cdsads.u-strasbg.fr/abs/2011ApJ...729...36K} {729, 36}

\bibitem[\protect\citeauthoryear{{Krumholz} \& {Kruijssen}}{{Krumholz} \&
  {Kruijssen}}{2015}]{Krumholz+15a}
{Krumholz} M.~R.,  {Kruijssen} J.~M.~D.,  2015, \mn@doi [\mnras]
  {10.1093/mnras/stv1670}, \href
  {http://cdsads.u-strasbg.fr/abs/2015MNRAS.453..739K} {453, 739}

\bibitem[\protect\citeauthoryear{{Krumholz} \& {Tan}}{{Krumholz} \&
  {Tan}}{2007}]{Krumholz&Tan07}
{Krumholz} M.~R.,  {Tan} J.~C.,  2007, \mn@doi [\apj] {10.1086/509101}, \href
  {http://cdsads.u-strasbg.fr/abs/2007ApJ...654..304K} {654, 304}

\bibitem[\protect\citeauthoryear{{Krumholz}, {Dekel}  \& {McKee}}{{Krumholz}
  et~al.}{2012}]{Krumholz+12}
{Krumholz} M.~R.,  {Dekel} A.,   {McKee} C.~F.,  2012, \mn@doi [\apj]
  {10.1088/0004-637X/745/1/69}, \href
  {http://adsabs.harvard.edu/abs/2012ApJ...745...69K} {745, 69}

\bibitem[\protect\citeauthoryear{{Krumholz}, {Fumagalli}, {da Silva}, {Rendahl}
   \& {Parra}}{{Krumholz} et~al.}{2015}]{Krumholz+15b}
{Krumholz} M.~R.,  {Fumagalli} M.,  {da Silva} R.~L.,  {Rendahl} T.,   {Parra}
  J.,  2015, \mn@doi [\mnras] {10.1093/mnras/stv1374}, \href
  {http://cdsads.u-strasbg.fr/abs/2015MNRAS.452.1447K} {452, 1447}

\bibitem[\protect\citeauthoryear{{Krumholz}, {Kruijssen}  \&
  {Crocker}}{{Krumholz} et~al.}{2017}]{Krumholz+17}
{Krumholz} M.~R.,  {Kruijssen} J.~M.~D.,   {Crocker} R.~M.,  2017, \mn@doi
  [\mnras] {10.1093/mnras/stw3195}, \href
  {http://cdsads.u-strasbg.fr/abs/2017MNRAS.466.1213K} {466, 1213}

\bibitem[\protect\citeauthoryear{{Krumholz}, {McKee}  \&
  {Bland-Hawthorn}}{{Krumholz} et~al.}{2019}]{Krumholz+19}
{Krumholz} M.~R.,  {McKee} C.~F.,   {Bland-Hawthorn} J.,  2019, \araa, \href
  {https://ui.adsabs.harvard.edu/\#abs/2018arXiv181201615K} {}

\bibitem[\protect\citeauthoryear{{Launhardt}, {Zylka}  \& {Mezger}}{{Launhardt}
  et~al.}{2002}]{Launhardt+02}
{Launhardt} R.,  {Zylka} R.,   {Mezger} P.~G.,  2002, \mn@doi [\aap]
  {10.1051/0004-6361:20020017}, \href
  {http://adsabs.harvard.edu/abs/2002A%26A...384..112L} {384, 112}

\bibitem[\protect\citeauthoryear{{Lee}, {Miville-Desch{\^e}nes}  \&
  {Murray}}{{Lee} et~al.}{2016}]{Lee+16}
{Lee} E.~J.,  {Miville-Desch{\^e}nes} M.-A.,   {Murray} N.~W.,  2016, \mn@doi
  [\apj] {10.3847/1538-4357/833/2/229}, \href
  {http://adsabs.harvard.edu/abs/2016ApJ...833..229L} {833, 229}

\bibitem[\protect\citeauthoryear{{Leitherer} et~al.,}{{Leitherer}
  et~al.}{1999}]{Leitherer+99}
{Leitherer} C.,  et~al., 1999, \mn@doi [\apjs] {10.1086/313233}, \href
  {http://adsabs.harvard.edu/cgi-bin/nph-bib\_query?bibcode=1999ApJS..123....3L\&db\_key=AST}
  {123, 3}

\bibitem[\protect\citeauthoryear{{Leroy} et~al.,}{{Leroy}
  et~al.}{2013}]{Leroy+13}
{Leroy} A.~K.,  et~al., 2013, \mn@doi [\aj] {10.1088/0004-6256/146/2/19}, \href
  {http://cdsads.u-strasbg.fr/abs/2013AJ....146...19L} {146, 19}

\bibitem[\protect\citeauthoryear{{Lesch}, {Biermann}, {Crusius}, {Reuter},
  {Dahlem}, {Barteldrees}  \& {Wielebinski}}{{Lesch} et~al.}{1990}]{Lesch+90}
{Lesch} H.,  {Biermann} P.~L.,  {Crusius} A.,  {Reuter} H.~P.,  {Dahlem} M.,
  {Barteldrees} A.,   {Wielebinski} R.,  1990, \mn@doi [\mnras]
  {10.1093/mnras/242.2.194}, \href
  {https://ui.adsabs.harvard.edu/abs/1990MNRAS.242..194L} {242, 194}

\bibitem[\protect\citeauthoryear{{Li}, {Gerhard}, {Shen}, {Portail}  \&
  {Wegg}}{{Li} et~al.}{2016}]{Li+16}
{Li} Z.,  {Gerhard} O.,  {Shen} J.,  {Portail} M.,   {Wegg} C.,  2016, \mn@doi
  [\apj] {10.3847/0004-637X/824/1/13}, \href
  {http://cdsads.u-strasbg.fr/abs/2016ApJ...824...13L} {824, 13}

\bibitem[\protect\citeauthoryear{{Li}, {Bryan}  \& {Ostriker}}{{Li}
  et~al.}{2017}]{Li+17}
{Li} M.,  {Bryan} G.~L.,   {Ostriker} J.~P.,  2017, \mn@doi [\apj]
  {10.3847/1538-4357/aa7263}, \href
  {http://cdsads.u-strasbg.fr/abs/2017ApJ...841..101L} {841, 101}

\bibitem[\protect\citeauthoryear{{Longmore} et~al.,}{{Longmore}
  et~al.}{2013}]{Longmore+13a}
{Longmore} S.~N.,  et~al., 2013, \mn@doi [\mnras] {10.1093/mnras/sts376}, \href
  {http://cdsads.u-strasbg.fr/abs/2013MNRAS.429..987L} {429, 987}

\bibitem[\protect\citeauthoryear{{Martizzi}, {Fielding}, {Faucher-Gigu{\`e}re}
  \& {Quataert}}{{Martizzi} et~al.}{2016}]{Martizzi+16}
{Martizzi} D.,  {Fielding} D.,  {Faucher-Gigu{\`e}re} C.-A.,   {Quataert} E.,
  2016, \mn@doi [\mnras] {10.1093/mnras/stw745}, \href
  {http://cdsads.u-strasbg.fr/abs/2016MNRAS.459.2311M} {459, 2311}

\bibitem[\protect\citeauthoryear{{McClure-Griffiths}, {Green}, {Hill},
  {Lockman}, {Dickey}, {Gaensler}  \& {Green}}{{McClure-Griffiths}
  et~al.}{2013}]{McClure-Griffiths+13}
{McClure-Griffiths} N.~M.,  {Green} J.~A.,  {Hill} A.~S.,  {Lockman} F.~J.,
  {Dickey} J.~M.,  {Gaensler} B.~M.,   {Green} A.~J.,  2013, \mn@doi [\apjl]
  {10.1088/2041-8205/770/1/L4}, \href
  {http://cdsads.u-strasbg.fr/abs/2013ApJ...770L...4M} {770, L4}

\bibitem[\protect\citeauthoryear{{McMillan}}{{McMillan}}{2017}]{McMillan17}
{McMillan} P.~J.,  2017, \mn@doi [\mnras] {10.1093/mnras/stw2759}, \href
  {http://cdsads.u-strasbg.fr/abs/2017MNRAS.465...76M} {465, 76}

\bibitem[\protect\citeauthoryear{{Miller} \& {Bregman}}{{Miller} \&
  {Bregman}}{2015}]{Miller&Bregman15}
{Miller} M.~J.,  {Bregman} J.~N.,  2015, \mn@doi [\apj]
  {10.1088/0004-637X/800/1/14}, \href
  {http://cdsads.u-strasbg.fr/abs/2015ApJ...800...14M} {800, 14}

\bibitem[\protect\citeauthoryear{{Miller} \& {Bregman}}{{Miller} \&
  {Bregman}}{2016}]{Miller&Bregman16}
{Miller} M.~J.,  {Bregman} J.~N.,  2016, \mn@doi [\apj]
  {10.3847/0004-637X/829/1/9}, \href
  {http://cdsads.u-strasbg.fr/abs/2016ApJ...829....9M} {829, 9}

\bibitem[\protect\citeauthoryear{{Molinari} et~al.,}{{Molinari}
  et~al.}{2011}]{Molinari+11}
{Molinari} S.,  et~al., 2011, \mn@doi [\apjl] {10.1088/2041-8205/735/2/L33},
  \href {http://cdsads.u-strasbg.fr/abs/2011ApJ...735L..33M} {735, L33}

\bibitem[\protect\citeauthoryear{{Morris} \& {Serabyn}}{{Morris} \&
  {Serabyn}}{1996}]{Morris&Serabyn96}
{Morris} M.,  {Serabyn} E.,  1996, \mn@doi [\araa]
  {10.1146/annurev.astro.34.1.645}, \href
  {http://cdsads.u-strasbg.fr/abs/1996ARA%26A..34..645M} {34, 645}

\bibitem[\protect\citeauthoryear{{Navarro}, {Frenk}  \& {White}}{{Navarro}
  et~al.}{1996}]{Navarro+96}
{Navarro} J.~F.,  {Frenk} C.~S.,   {White} S.~D.~M.,  1996, \mn@doi [\apj]
  {10.1086/177173}, \href {http://cdsads.u-strasbg.fr/abs/1996ApJ...462..563N}
  {462, 563}

\bibitem[\protect\citeauthoryear{{Ridley}, {Sormani}, {Tre{\ss}}, {Magorrian}
  \& {Klessen}}{{Ridley} et~al.}{2017}]{Ridley+17}
{Ridley} M.~G.~L.,  {Sormani} M.~C.,  {Tre{\ss}} R.~G.,  {Magorrian} J.,
  {Klessen} R.~S.,  2017, \mn@doi [\mnras] {10.1093/mnras/stx944}, \href
  {http://cdsads.u-strasbg.fr/abs/2017MNRAS.469.2251R} {469, 2251}

\bibitem[\protect\citeauthoryear{{Sarzi}, {Allard}, {Knapen}  \&
  {Mazzuca}}{{Sarzi} et~al.}{2007}]{Sarzi+07}
{Sarzi} M.,  {Allard} E.~L.,  {Knapen} J.~H.,   {Mazzuca} L.~M.,  2007, \mn@doi
  [\mnras] {10.1111/j.1365-2966.2007.12177.x}, \href
  {https://ui.adsabs.harvard.edu/abs/2007MNRAS.380..949S} {380, 949}

\bibitem[\protect\citeauthoryear{{Schmidt}}{{Schmidt}}{1959}]{Schmidt59}
{Schmidt} M.,  1959, \mn@doi [\apj] {10.1086/146614}, \href
  {http://cdsads.u-strasbg.fr/abs/1959ApJ...129..243S} {129, 243}

\bibitem[\protect\citeauthoryear{{Sembach} et~al.,}{{Sembach}
  et~al.}{2003}]{Sembach+03}
{Sembach} K.~R.,  et~al., 2003, \mn@doi [\apjs] {10.1086/346231}, \href
  {http://cdsads.u-strasbg.fr/abs/2003ApJS..146..165S} {146, 165}

\bibitem[\protect\citeauthoryear{{Seo}, {Kim}, {Kwak}, {Hsieh}, {Han}  \&
  {Hopkins}}{{Seo} et~al.}{2019}]{Seo+19}
{Seo} W.-Y.,  {Kim} W.-T.,  {Kwak} S.,  {Hsieh} P.-Y.,  {Han} C.,   {Hopkins}
  P.~F.,  2019, \mn@doi [\apj] {10.3847/1538-4357/aafc5f}, \href
  {http://adsabs.harvard.edu/abs/2019ApJ...872....5S} {872, 5}

\bibitem[\protect\citeauthoryear{{Smith} et~al.,}{{Smith}
  et~al.}{2017}]{Smith+17}
{Smith} B.~D.,  et~al., 2017, \mn@doi [\mnras] {10.1093/mnras/stw3291}, \href
  {http://cdsads.u-strasbg.fr/abs/2017MNRAS.466.2217S} {466, 2217}

\bibitem[\protect\citeauthoryear{{Sormani} \& {Barnes}}{{Sormani} \&
  {Barnes}}{2019}]{Sormani+19}
{Sormani} M.~C.,  {Barnes} A.~T.,  2019, \mn@doi [\mnras]
  {10.1093/mnras/stz046}, \href
  {http://cdsads.u-strasbg.fr/abs/2019MNRAS.484.1213S} {484, 1213}

\bibitem[\protect\citeauthoryear{{Sormani}, {Binney}  \& {Magorrian}}{{Sormani}
  et~al.}{2015}]{Sormani+15}
{Sormani} M.~C.,  {Binney} J.,   {Magorrian} J.,  2015, \mn@doi [\mnras]
  {10.1093/mnras/stv441}, \href
  {http://cdsads.u-strasbg.fr/abs/2015MNRAS.449.2421S} {449, 2421}

\bibitem[\protect\citeauthoryear{{Sormani}, {Tre{\ss}}, {Ridley}, {Glover},
  {Klessen}, {Binney}, {Magorrian}  \& {Smith}}{{Sormani}
  et~al.}{2018a}]{Sormani+17}
{Sormani} M.~C.,  {Tre{\ss}} R.~G.,  {Ridley} M.,  {Glover} S.~C.~O.,
  {Klessen} R.~S.,  {Binney} J.,  {Magorrian} J.,   {Smith} R.,  2018a, \mn@doi
  [\mnras] {10.1093/mnras/stx3258}, \href
  {http://cdsads.u-strasbg.fr/abs/2018MNRAS.475.2383S} {475, 2383}

\bibitem[\protect\citeauthoryear{{Sormani}, {Sobacchi}, {Fragkoudi}, {Ridley},
  {Tre{\ss}}, {Glover}  \& {Klessen}}{{Sormani} et~al.}{2018b}]{Sormani+18}
{Sormani} M.~C.,  {Sobacchi} E.,  {Fragkoudi} F.,  {Ridley} M.,  {Tre{\ss}}
  R.~G.,  {Glover} S.~C.~O.,   {Klessen} R.~S.,  2018b, \mn@doi [\mnras]
  {10.1093/mnras/sty2246}, \href
  {http://cdsads.u-strasbg.fr/abs/2018MNRAS.481....2S} {481, 2}

\bibitem[\protect\citeauthoryear{{Sormani} et~al.,}{{Sormani}
  et~al.}{2019}]{Sormani+19b}
{Sormani} M.~C.,  et~al., 2019, \mn@doi [\mnras] {10.1093/mnras/stz2054}, \href
  {https://ui.adsabs.harvard.edu/abs/2019MNRAS.488.4663S} {488, 4663}

\bibitem[\protect\citeauthoryear{{Springel}}{{Springel}}{2005}]{Springel05}
{Springel} V.,  2005, \mn@doi [\mnras] {10.1111/j.1365-2966.2005.09655.x},
  \href {http://cdsads.u-strasbg.fr/abs/2005MNRAS.364.1105S} {364, 1105}

\bibitem[\protect\citeauthoryear{{Su}, {Slatyer}  \& {Finkbeiner}}{{Su}
  et~al.}{2010}]{Su+10}
{Su} M.,  {Slatyer} T.~R.,   {Finkbeiner} D.~P.,  2010, \mn@doi [\apj]
  {10.1088/0004-637X/724/2/1044}, \href
  {http://cdsads.u-strasbg.fr/abs/2010ApJ...724.1044S} {724, 1044}

\bibitem[\protect\citeauthoryear{{Sukhbold}, {Ertl}, {Woosley}, {Brown}  \&
  {Janka}}{{Sukhbold} et~al.}{2016}]{Sukhbold+16}
{Sukhbold} T.,  {Ertl} T.,  {Woosley} S.~E.,  {Brown} J.~M.,   {Janka} H.-T.,
  2016, \mn@doi [\apj] {10.3847/0004-637X/821/1/38}, \href
  {http://adsabs.harvard.edu/abs/2016ApJ...821...38S} {821, 38}

\bibitem[\protect\citeauthoryear{{Tasker} \& {Bryan}}{{Tasker} \&
  {Bryan}}{2008}]{Tasker&Bryan08}
{Tasker} E.~J.,  {Bryan} G.~L.,  2008, \mn@doi [\apj] {10.1086/523889}, \href
  {https://ui.adsabs.harvard.edu/abs/2008ApJ...673..810T} {673, 810}

\bibitem[\protect\citeauthoryear{{Torrey}, {Hopkins}, {Faucher-Gigu{\`e}re},
  {Vogelsberger}, {Quataert}, {Kere{\v s}}  \& {Murray}}{{Torrey}
  et~al.}{2017}]{Torrey+17}
{Torrey} P.,  {Hopkins} P.~F.,  {Faucher-Gigu{\`e}re} C.-A.,  {Vogelsberger}
  M.,  {Quataert} E.,  {Kere{\v s}} D.,   {Murray} N.,  2017, \mn@doi [\mnras]
  {10.1093/mnras/stx254}, \href
  {http://cdsads.u-strasbg.fr/abs/2017MNRAS.467.2301T} {467, 2301}

\bibitem[\protect\citeauthoryear{{Utomo} et~al.,}{{Utomo}
  et~al.}{2017}]{Utomo+17}
{Utomo} D.,  et~al., 2017, \mn@doi [\apj] {10.3847/1538-4357/aa88c0}, \href
  {http://cdsads.u-strasbg.fr/abs/2017ApJ...849...26U} {849, 26}

\bibitem[\protect\citeauthoryear{{Utomo} et~al.,}{{Utomo}
  et~al.}{2018}]{Utomo+18}
{Utomo} D.,  et~al., 2018, \mn@doi [\apj] {10.3847/2041-8213/aacf8f}, \href
  {https://ui.adsabs.harvard.edu/#abs/2018ApJ...861L..18U} {861, L18}

\bibitem[\protect\citeauthoryear{{Vutisalchavakul}, {Evans}  \&
  {Heyer}}{{Vutisalchavakul} et~al.}{2016}]{Vutisalchavakul+16}
{Vutisalchavakul} N.,  {Evans} II N.~J.,   {Heyer} M.,  2016, \mn@doi [\apj]
  {10.3847/0004-637X/831/1/73}, \href
  {http://cdsads.u-strasbg.fr/abs/2016ApJ...831...73V} {831, 73}

\bibitem[\protect\citeauthoryear{{Walch} \& {Naab}}{{Walch} \&
  {Naab}}{2015}]{Walch&Naab15}
{Walch} S.,  {Naab} T.,  2015, \mn@doi [\mnras] {10.1093/mnras/stv1155}, \href
  {http://cdsads.u-strasbg.fr/abs/2015MNRAS.451.2757W} {451, 2757}

\bibitem[\protect\citeauthoryear{{Wegg} \& {Gerhard}}{{Wegg} \&
  {Gerhard}}{2013}]{Wegg&Gerhard13}
{Wegg} C.,  {Gerhard} O.,  2013, \mn@doi [\mnras] {10.1093/mnras/stt1376},
  \href {http://cdsads.u-strasbg.fr/abs/2013MNRAS.435.1874W} {435, 1874}

\bibitem[\protect\citeauthoryear{{Wegg}, {Gerhard}  \& {Portail}}{{Wegg}
  et~al.}{2015}]{Wegg+15}
{Wegg} C.,  {Gerhard} O.,   {Portail} M.,  2015, \mn@doi [\mnras]
  {10.1093/mnras/stv745}, \href
  {http://cdsads.u-strasbg.fr/abs/2015MNRAS.450.4050W} {450, 4050}

\bibitem[\protect\citeauthoryear{{Wolfire}, {McKee}, {Hollenbach}  \&
  {Tielens}}{{Wolfire} et~al.}{2003}]{Wolfire+03}
{Wolfire} M.~G.,  {McKee} C.~F.,  {Hollenbach} D.,   {Tielens} A.~G.~G.~M.,
  2003, \mn@doi [\apj] {10.1086/368016}, \href
  {https://ui.adsabs.harvard.edu/abs/2003ApJ...587..278W} {587, 278}

\bibitem[\protect\citeauthoryear{{Yusef-Zadeh} et~al.,}{{Yusef-Zadeh}
  et~al.}{2009}]{Yusef-Zadeh+09}
{Yusef-Zadeh} F.,  et~al., 2009, \mn@doi [\apj] {10.1088/0004-637X/702/1/178},
  \href {http://cdsads.u-strasbg.fr/abs/2009ApJ...702..178Y} {702, 178}

\bibitem[\protect\citeauthoryear{{da Silva}, {Fumagalli}  \& {Krumholz}}{{da
  Silva} et~al.}{2012}]{daSilva+12}
{da Silva} R.~L.,  {Fumagalli} M.,   {Krumholz} M.,  2012, \mn@doi [\apj]
  {10.1088/0004-637X/745/2/145}, \href
  {http://cdsads.u-strasbg.fr/abs/2012ApJ...745..145D} {745, 145}

\makeatother
\end{thebibliography}
    

\appendix
\section{Sensitivity to numerical parameters}\label{appendix}
\label{append}

\begin{figure*}
\includegraphics[width=\textwidth]{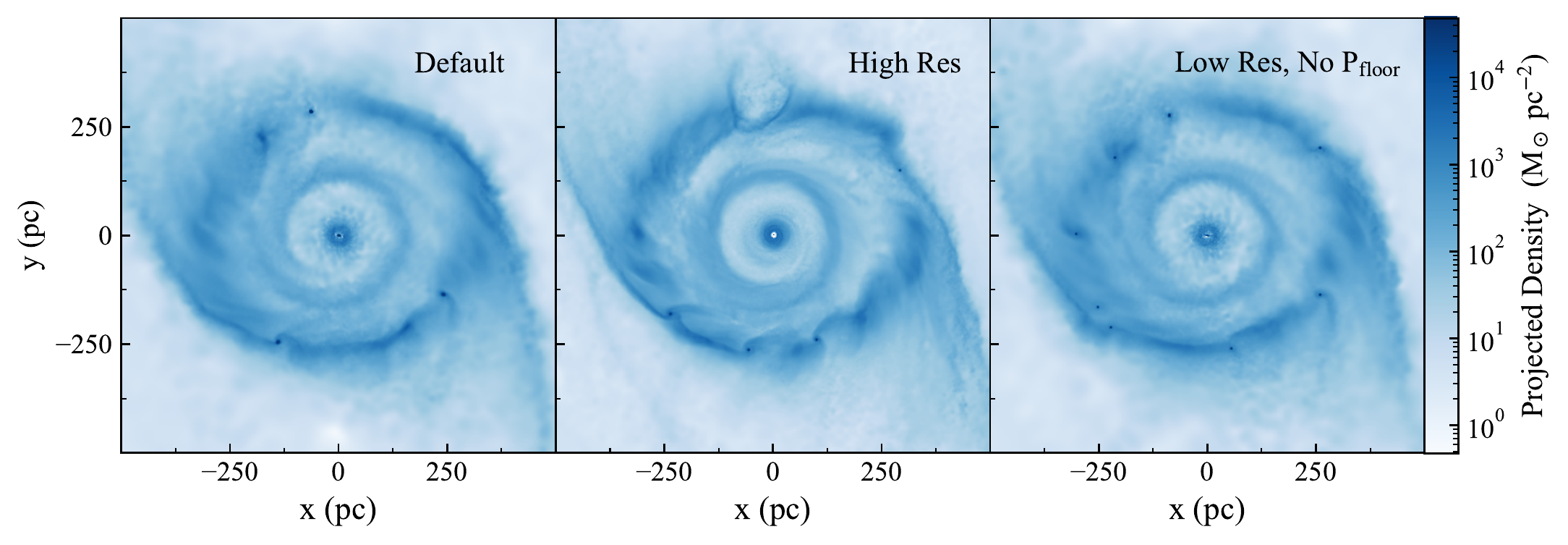}
\caption{Face-on gas density projection in the central 500~pc of the Galaxy for the default simulation (\textit{left panel}), high resolution simulation (\textit{middle panel)} and low resolution simulation in the absence of pressure floor (\textit{right panel}). The snapshots have been taken at $t = 520$~Myr.}
\label{CMZ_comparison}
\end{figure*}

\begin{figure}
\includegraphics[width=0.49\textwidth]{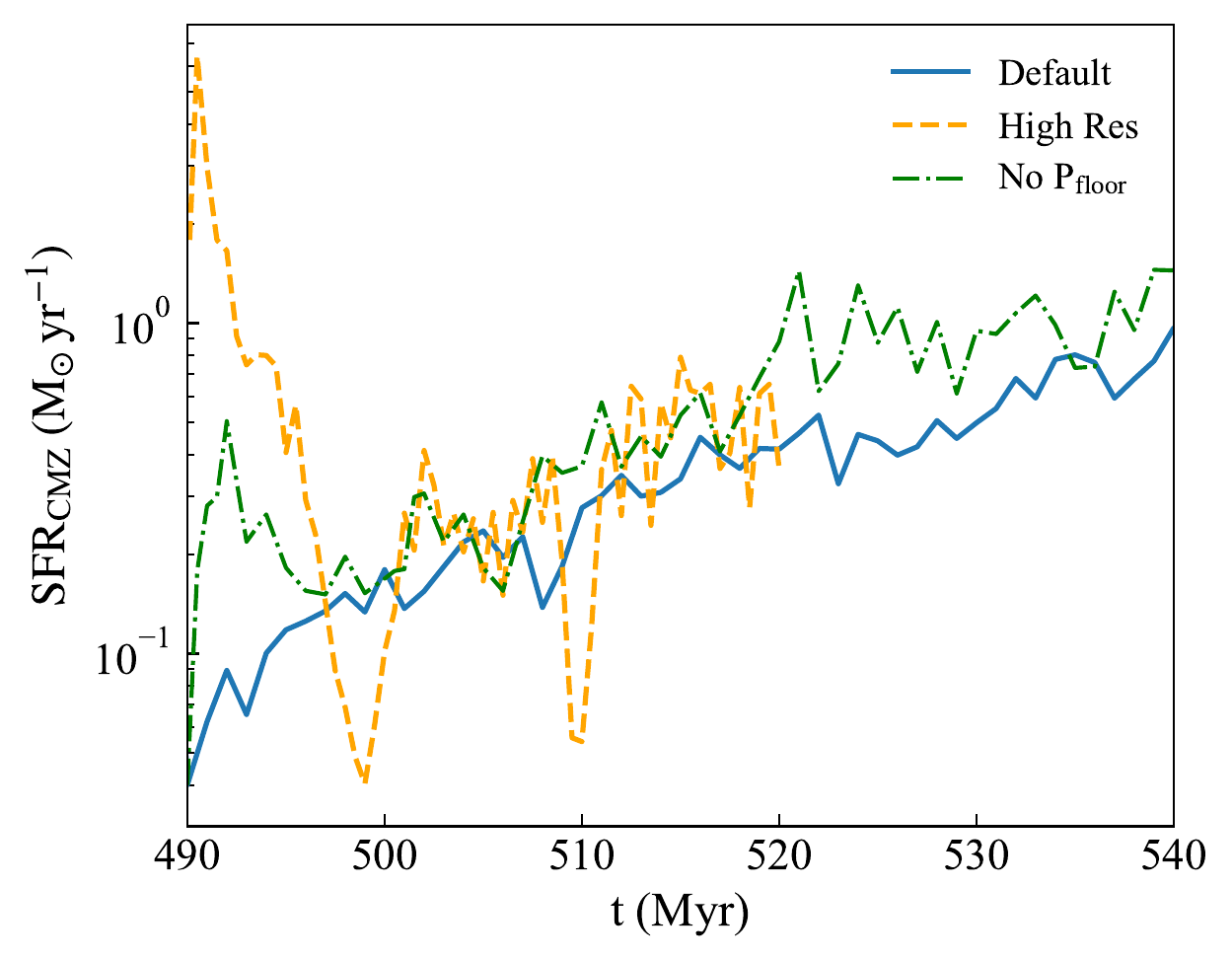}
\caption{SFR versus time in the default simulation from the main text (\textit{blue solid line}), high resolution simulation (\textit{orange dashed line)} and simulation in the absence of pressure floor (\textit{green dash-dotted line}).}
\label{SFR_comparison}
\end{figure}

\begin{figure}
\includegraphics[width=0.49\textwidth]{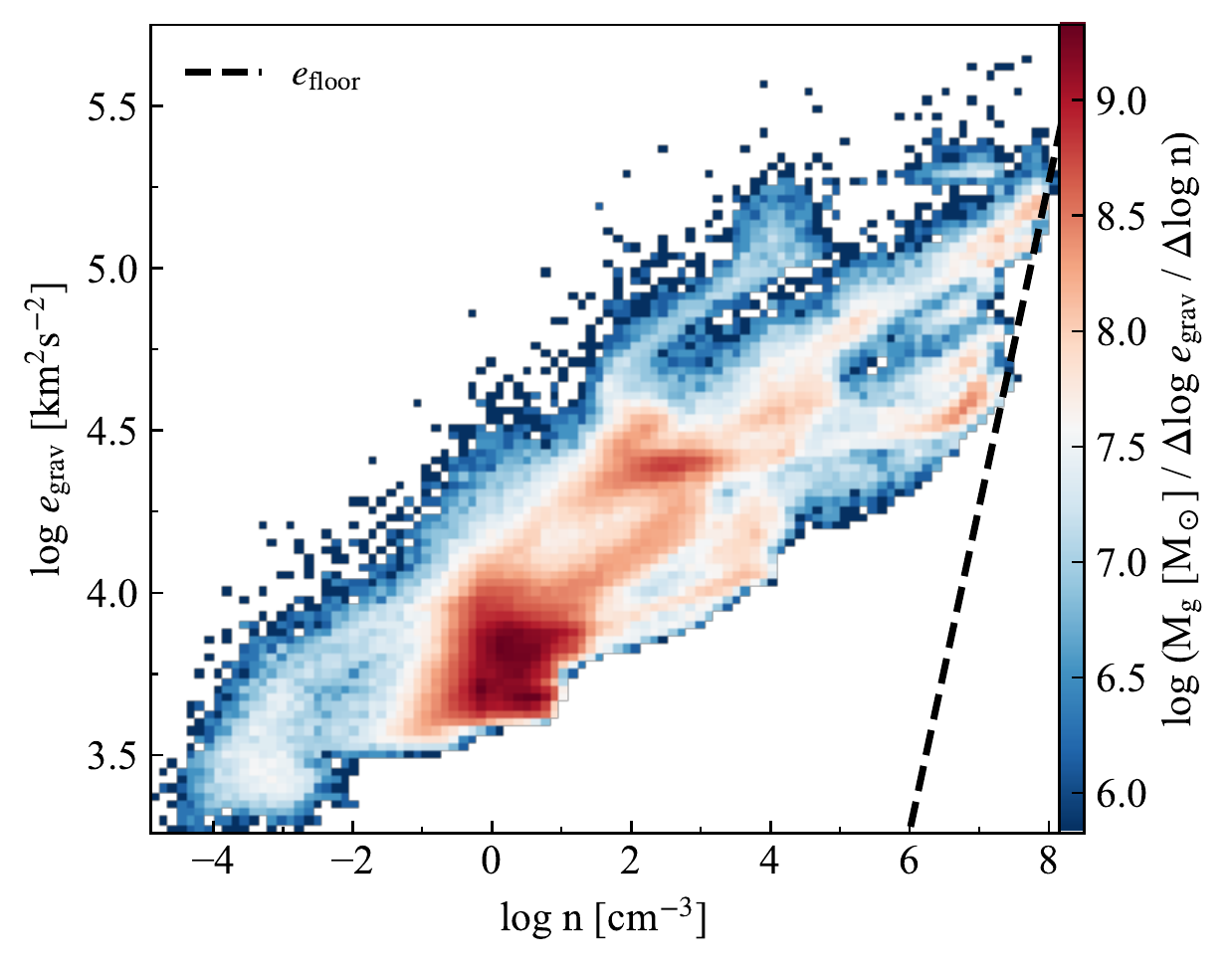}
\caption{Two-dimensional distribution of gas mass as a function of volume density and specific gravitational energy, $e_\mathrm{grav}$, in logarithmically-spaced bins. The black dashed line indicates the specific internal energy supplied by the artificial pressure, $e_\mathrm{floor}$. The data have been extracted from the simulation at low resolution and without pressure floor at $t=520$~Myr. }
\label{Pfloor}
\end{figure}

In this Appendix, we conduct studies to verify the robustness of our results to two numerical effects: resolution and pressure floor. To this end, we run two additional simulations starting with initial conditions extracted from the fiducial simulation discussed in the main text at $t=490$~Myr, when the CMZ is at a minimum of its star formation activity, and thus closest to the conditions we associate with the present-day Milky Way. The first of our studies uses a mass resolution and softening length of $200\,\mo$ and 0.1~pc, respectively; this is a factor of 10 smaller mass and softening length than our standard case. We run this high resolution case for 30 Myr, which is several orbits of the CMZ. The second simulation is run at the same resolution as the default simulation, but without imposing an artificial pressure support (see \autoref{Cooling and Feedback}), and it is evolved for 50 Myr. The pressure floor term is used in the default simulation to suppress gas fragmentation when the Jeans length is lower than the spatial resolution, but at the potential price of inhibiting gravitational collapse and star formation in regions where it is imposed.

\autoref{CMZ_comparison} compares the gas surface density in the central 500~pc region of the Galaxy for the default simulation (left panel), high resolution simulation (middle panel) and no pressure floor simulation (right panel). We find no significant difference in the global CMZ morphology between the three cases. The ring shape is nearly identical except for the hole at $x=0$~pc and $y=250$~pc in the high resolution simulation. This hole is the result of a superbubble explosion. In the lower resolution runs, the superbubble region is occupied by a cloud. This difference may be an indication that feedback is more effective at higher resolution, or it may simply be a matter of timing -- the superbubble may form slightly early in the high resolution simulation than in the lower resolution cases. A second minor difference is the number of molecular clouds: the default simulation yields a slightly smaller number of somewhat more massive clouds, due to the combination of resolution effects and suppression of fragmentation by the pressure floor. However, this difference is clearly only significant on small scales.

\autoref{SFR_comparison} shows the SFR as a function of time in the three simulations. Both the no pressure floor and the higher resolution case experience a brief burst of star formation at first, as the removal of the pressure floor / reduction in softening length allow a sudden increase in gas density. However, after a few Myr stellar feedback pushes the systems back towards equilibrium. From that point on, the SFR as a function of time is very similar in the three simulations. The SFR in the high resolution case fluctuates more rapidly, which is not surprising given its improved resolution, but has the same time-averaged trend. The SFR in the simulation without a pressure floor is systematically above the SFR in the default simulation, but only by tens of percent. Clearly neither the resolution nor the pressure floor play a significant role in dictating the overall SFR.

We can further understand why the pressure floor is so unimportant to the overall dynamics by directly examining the energetics in the problem. The pressure floor is negligible for gas at $T\sim 10^4$~K, because the thermal pressure is always larger than the floor pressure unless $n>10^5$~cm$^{-3}$. The pressure floor does significantly increase the thermal energy for gas at $T \sim 10$~K. However, this has very little influence on the overall rate of collapse or star formation, because for this gas the thermal energy is energetically unimportant compared to gravity. To demonstrate this point, in \autoref{Pfloor} we show the two-dimensional distribution of gas mass as a function of number density and specific gravitational energy, $e_\mathrm{grav}$. The data have been extracted by the simulation without pressure floor, to avoid contamination due to the presence of an artificial pressure support, and the gravitational energy includes only the self-gravitational energy of the gas (computed in a gauge where $e_{\rm grav}\to 0$ as particles move off to infinity), not any contribution from the background stellar / dark matter potential. The black dashed line shows the specific internal energy supplied by the artificial pressure, $e_\mathrm{floor}= N_\mathrm{J}^2 G \rho_\mathrm{g} \,\mathrm{max}(h_\mathrm{g},s)^2/\gamma/(\gamma-1)$ (see \autoref{Methods} for the meaning of all these terms). We see that the pressure floor contribution is comparable to the gravitational specific energy only at densities $n\gtrsim 10^6$~cm$^{-3}$. Therefore, the thermal energy required to suppress the gravitational collapse is supplied by the pressure-floor support at very high densities only. Gas at $n\lesssim 10^6$~cm$^{-3}$ may acquire extra thermal energy from the pressure floor, but this does not inhibit collapse because the amount of energy added is very small. This explains the small difference between the SFRs in the simulations with and without pressure floor.

\bsp	
\label{lastpage}
\end{document}